
\def\ignore#1{}
 

\newcount\sectnum
\newcount\subsectnum
\newcount\eqnumber

\global\eqnumber=1\sectnum=0


\def\lab{(\the\sectnum.\the\eqnumber)}



\def\show#1{#1}



\def\smskip{\vskip 5 pt}
\def\medskip{\vskip 10 pt}
\def\bigskip{\vskip 15 pt}
\def\pn{\par\noindent}

\def\lf{\left}
\def\ri{\right}

\def\argmin{\mathop{\arg \min}}

\def\implies{\Rightarrow}

\def\frac#1#2{{#1\over #2}}

\def\ol#1{\overline{#1}}

\def\l{\lambda}
\def\g{\gamma}
\def\m{\mu}
\def\p{\pi}

\def\d{\delta}

\def\re{\Re}

\def\tl{\tilde}

\def\old#1{}
\def\leaderfill{\leaders\hbox to 1em{\hss.\hss}\hfill}


\parindent=2pc
\baselineskip=15pt
\vsize=8.7 true in
\voffset=0.125 true in
\parskip=3pt


\def\minprob#1#2#3{$$\eqalign{&\hbox{minimize\ \ }#1\cr &\hbox{subject to\ \
}#2\cr}\ifnum 0=#3{}\else\eqno(#3)\fi$$}        
     
\def\maxprob#1#2#3{$$\eqalign{&\hbox{maximize\ \ }#1\cr &\hbox{subject to\ \
}#2\cr}\ifnum 0=#3{}\else\eqno(#3)\fi$$}        
     
\def\aligntwo#1#2#3#4#5{$$\eqalign{#1&#2\cr #3&#4\cr}
\ifnum 0=#5{}\else\eqno(#5)\fi$$}
\def\alignthree#1#2#3#4#5#6#7{$$\eqalign{#1&#2\cr #3&#4\cr #5&#6\cr}
\ifnum 0=#7{}\else\eqno(#7)\fi$$}


\def\eqnum{\eqno{\hbox{(\the\sectnum.\the\eqnumber)}\global\advance\eqnumber
by1}}

\def\eqnu{\eqno{\hbox{(\the\sectnum.\the\eqnumber)}\global\advance\eqnumber
by1}}

\newcount\examplnumber
\def\examplnum{\global\advance\examplnumber by1}

\newcount\figrnumber
\def\figrnum{\global\advance\figrnumber by1}

\newcount\propnumber
\def\propnum{\global\advance\propnumber by1}

\newcount\defnumber
\def\defnum{\global\advance\defnumber by1}

\newcount\lemmanumber
\def\lemmanum{\global\advance\lemmanumber by1}

\newcount\assumptionnumber
\def\assumptionnum{\global\advance\assumptionnumber by1}

\newcount\conditionnumber
\def\conditionnum{\global\advance\conditionnumber by1}

\def\exampl{\the\sectnum.\the\examplnumber}
\def\figr{\the\sectnum.\the\figrnumber}
\def\propn{\the\sectnum.\the\propnumber}
\def\defn{\the\sectnum.\the\defnumber}
\def\lemman{\the\sectnum.\the\lemmanumber}
\def\assumptionn{\the\sectnum.\the\assumptionnumber}
\def\condn{\the\sectnum.\the\conditionnumber}

\def\section#1{\goodbreak\vskip 3pc plus 6pt minus 3pt\leftskip=-2pc
   \global\advance\sectnum by 1\eqnumber=1
\global\examplnumber=1\figrnumber=1\propnumber=1\defnumber=1\lemmanumber=1\assumptionnumber=1 \conditionnumber =1%
   \line{\hfuzz=1pc{\hbox to 3pc{\bf 
   \vtop{\hfuzz=1pc\hsize=38pc\hyphenpenalty=10000\noindent\uppercase{\the\sectnum.\quad #1}}\hss}}
			\hfill}
			\leftskip=0pc\nobreak\tenf
			\vskip 1pc plus 4pt minus 2pt\noindent\ignorespaces}



\def\sect#1{\noindent\leftskip=-2pc\tenf
   \goodbreak\vskip 1pc plus 4pt minus 2pt
                \global\advance\subsectnum by 1\eqnumber=1
   \line{\hfuzz=1pc{\hbox to 3pc{\bf 
   \vtop{\hfuzz=1pc\hsize=38pc\hyphenpenalty=10000\noindent\uppercase{{\bf #1}}}\hss}}
                        \hfill}
   \leftskip=0pc\nobreak\tenf
                        \vskip 1pc plus 4pt minus 2pt\nobreak\noindent\ignorespaces}

\def\subsection#1{\noindent\leftskip=0pc\tenf
   \goodbreak\vskip 1pc plus 4pt minus 2pt
   \line{\hfuzz=1pc{\hbox to 3pc{\bf 
   \vtop{\hfuzz=1pc\hsize=38pc\hyphenpenalty=10000\noindent{\bf #1}}\hss}}
                        \hfill}
   \leftskip=0pc\nobreak\tenf
                        \vskip 1pc plus 4pt minus 2pt\nobreak\noindent\ignorespaces}
\def\subsubsection#1{\goodbreak\vskip 1pc plus 4pt minus 2pt
   \hfuzz=3pc\leftskip=0pc\noindent\tenit #1 \nobreak\tenf\vskip 6pt plus 1pt
                                minus 1pt\nobreak\ignorespaces\leftskip=0pc}
%

\def\beginexample#1{\noindent\goodbreak\vskip 6pt plus 1pt minus 1pt
\noindent
  \hbox {\bf Example #1\hss}
  \nobreak\vskip 4pt plus 1pt minus 1pt \nobreak\noindent\ninef
  \global\advance
                \leftskip by\parindent\pn}
\def\endexample{\vskip 12pt\tenf\par
  \global\advance\leftskip by -\parindent
  }

\def\beginexercise#1{\noindent\goodbreak\vskip 6pt plus 1pt minus 1pt \noindent\global\normalbaselineskip=12pt
  \hbox {\bf Exercise #1\hss}
  \nobreak\vskip 4pt plus 1pt minus 1pt 
  \nobreak\noindent\ninef\global\advance\leftskip
                        by\parindent\pn}
\def\endexercise{\vskip 12pt\tenf\par
  \global\advance\leftskip by -\parindent
  }

\def\beginsection#1{\noindent\goodbreak\vskip 6pt plus 1pt minus 1pt \noindent\global\normalbaselineskip=12pt
  \hbox {\it #1\hss}
  \vskip 0.1pt plus 1pt minus 1pt \nobreak\noindent\ninef\global\advance
                \leftskip by\parindent\noindent\pn}
\def\endsection{\vskip 12pt\tenf\par
  \global\advance\leftskip by -\parindent
}

%


\def\proposition#1{\smskip\pn{\bf Proposition #1}\quad}
\def\proof{\smskip\pn{\bf Proof:}\quad} 
 
\def\assumption#1{\smskip\pn{\bf Assumption #1}\quad}

 \def\qed{\quad{\bf
Q.E.D.} \par\bigskip}
\def\ref{\smskip\pn}

\def\chapter#1#2{{\bf \centerline{\helbigbig
{#1}}}\bigskip\bigskip{\bf \centerline{\helbigbig
{#2}}}\bigskip\bigskip} 



\def\longpapertitle#1#2#3{{\bf \centerline{\helbigb
{#1}}}\bigskip{\bf \centerline{\helbigb
{#2}}}\bigskip\bigskip{\centerline{
by}}\bigskip{\bf \centerline{
{#3}}}\bigskip\bigskip} 


\def\nitem#1{\smskip\item{#1}}

\newcount\alphanum
\newcount\romnum

\def\alphaenumerate{\ifcase\alphanum \or (a)\or (b)\or (c)\or (d)\or (e)\or
(f)\or (g)\or (h)\or (i)\or (j)\or (k)\fi}
\def\romenumerate{\ifcase\romnum \or (i)\or (ii)\or (iii)\or (iv)\or (v)\or
(vi)\or (vii)\or (viii)\or (ix)\or (x)\or (xi)\fi}

\def\alist{\begingroup\vskip10pt\alphanum=1
\parskip=2pt\parindent=0pt \leftskip=3pc
\everypar{\llap{{\rm\alphaenumerate\hskip1em}}\advance\alphanum by1}}

\def\nolist{\begingroup\vskip10pt\alphanum=0
\parskip=2pt\parindent=0pt \leftskip=3pc
\everypar{\llap{\global\advance\alphanum by1(\the\alphanum)\hskip1em}}}

\def\romlist{\begingroup\vskip10pt\romnum=1
\parskip=2pt\parindent=0pt \leftskip=5pc
\everypar{\llap{{\rm\romenumerate\hskip1em}}\advance\romnum by1}}



\long\def\fig#1#2#3{\vbox{\vskip1pc\vskip#1
\prevdepth=12pt \baselineskip=12pt
\vskip1pc
\hbox to\hsize{\hfill\vtop{\hsize=25pc\noindent{\eightbf Figure #2\ }
{\eightpoint#3}}\hfill}}}

\long\def\widefig#1#2#3{\vbox{\vskip1pc\vskip#1
\prevdepth=12pt \baselineskip=12pt
\vskip1pc
\hbox to\hsize{\hfill\vtop{\hsize=28pc\noindent{\eightbf Figure #2\ }
{\eightpoint#3}}\hfill}}}

\long\def\table#1#2{\vbox{\vskip0.5pc
\prevdepth=12pt \baselineskip=12pt
\hbox to\hsize{\hfill\vtop{\hsize=25pc\noindent{\eightbf Table #1\ }
{\eightpoint#2}}\hfill}}}

 
\def\rightheadline#1{\headline{\tenrm\hfil #1}}


\long\def\leftfig#1#2{\vbox{\smskip\hsize=220pt
\vtop{{\noindent {\bf #1}}}
\smskip
\noindent
\vbox{{\noindent #2}}
}}

\long\def\rightfig#1#2#3{\vbox{\smskip\vskip#1
\prevdepth=12pt \baselineskip=12pt
\hsize=210pt
\smskip
\vbox{\noindent{\eightbold #2}
\hskip1em{\eightpoint#3}}
}}

\long\def\concept#1#2#3#4#5{\bigskip\hrule
\vbox{\hbox{\leftfig{#1}{#2} \hskip3em
\rightfig{#3}{#4}{#5}} \smskip}
\hrule\bigskip}


\long\def\bconcept#1#2#3#4#5#6#7{
\vbox{
\hbox to \hsize{\vtop{\par #1}}
\concept{#2}{#3}{#4}{#5}{#6}
\hbox to \hsize{\vtop{\par #7}}
\smskip}
}




\def\boxit#1{\vbox{\hrule\hbox{\vrule\kern3pt
                                \vbox{\kern3pt#1\kern3pt}\kern3pt\vrule}\hrule}}
\def\centerboxit#1{$$\vbox{\hrule\hbox{\vrule\kern3pt
                                \vbox{\kern3pt#1\kern3pt}\kern3pt\vrule}\hrule}$$}

\long\def\boxtext#1#2{$$\boxit{\vbox{\hsize #1\noindent\strut #2\strut}}$$}

%
%
%

\def\picture #1 by #2 (#3){
  \vbox to #2{
    \hrule width #1 height 0pt depth 0pt
    \vfill
    \special{picture #3} 
    }
  }

\def\scaledpicture #1 by #2 (#3 scaled #4){{
  \dimen0=#1 \dimen1=#2
  \divide\dimen0 by 1000 \multiply\dimen0 by #4
  \divide\dimen1 by 1000 \multiply\dimen1 by #4
  \picture \dimen0 by \dimen1 (#3 scaled #4)}
  }

%
%

\long\def\captfig#1#2#3#4#5{\vbox{\vskip1pc
\hbox to\hsize{\hfill{\picture #1 by #2 (#3)}\hfill}
\prevdepth=9pt \baselineskip=9pt
\vskip1pc
\hbox to\hsize{\hfill\vtop{\hsize=24pc\noindent{\eightbold Figure #4}
\hskip1em{\eightpoint#5}}\hfill}}}

%
%
%

\def\illustration #1 by #2 (#3){
  \vskip#2\hskip#1\special{illustration #3} 
    }

\def\scaledillustration #1 by #2 (#3 scaled #4){{
  \dimen0=#1 \dimen1=#2
  \divide\dimen0 by 1000 \multiply\dimen0 by #4
  \divide\dimen1 by 1000 \multiply\dimen1 by #4
  \illustration \dimen0 by \dimen1 (#3 scaled #4)}
  }


\newbox\graybox
\newdimen\xgrayspace
\newdimen\ygrayspace
%
%
%
%
%
%
%
%
%

\def\Textshade#1#2#3#4#5#6{%
    \xgrayspace=#4pt%
    \ygrayspace=#4pt%
    \def\grayshade{#3}%
    \def\linewidth{#5}%
    \def\theradius{#6}%
    \setbox\graybox=\hbox{\surroundboxa{#2}}%
    \hbox{%
    \hbox to 0pt{%
    \PScommands
}%
    \box\graybox}}%
%
%
\long%

\long%
\def\Parashade#1#2#3#4#5#6#7{%
    \xgrayspace=#4pt%
    \ygrayspace=#4pt%
    \def\grayshade{#3}%
    \def\linewidth{#5}%
    \def\theradius{#6}%
    \def\thevskip{#7pt}%
    \setbox\graybox=\hbox{\surroundboxb{#2}}%
    \vskip\thevskip%
    \hbox{%
    \hbox to 0pt{%
    \PScommands
}%
     \box\graybox}%
     \vskip\thevskip%
}%
%
%
%
\long\def\surroundboxa#1{\leavevmode\hbox{\vtop{%
\vbox{\kern\ygrayspace%
\hbox{\kern\xgrayspace#1%
      \kern\xgrayspace}}\kern\ygrayspace}}}
%
%
\long\def\surroundboxb#1{\leavevmode\hbox{\vtop{%
\vbox{\kern\ygrayspace%
\hbox{\kern\xgrayspace\vbox{\advance\hsize-2\xgrayspace#1}%
      \kern\xgrayspace}}\kern\ygrayspace}}}
%
%
%
\long\def\PScommands{%
\special{rawpostscript
/sharpbox{%
           newpath
           xmin ymin moveto
           xmin ymax lineto
           xmax ymax lineto
           xmax ymin lineto
           xmin ymin lineto
           closepath 
          }bind def
}%
\special{rawpostscript
/sharpboxnb{%
           newpath
           xmin ymin moveto
           xmin ymax lineto
           xmax ymax lineto
           xmax ymin lineto
          }bind def
}%
\special{rawpostscript
/sharpboxnt{%
           newpath
           xmin ymax moveto
           xmin ymin lineto
           xmax ymin lineto
           xmax ymax lineto
          }bind def
}%
\special{rawpostscript
/roundbox{%
           newpath
           xmin radius add ymin moveto
           xmax ymin xmax ymax radius arcto
           xmax ymax xmin ymax radius arcto
           xmin ymax xmin ymin radius arcto
           xmin ymin xmax ymin radius arcto 16 {pop} repeat
           closepath
          }bind def
}%
\special{rawpostscript
/sharpcorners{%
               sharpbox gsave grayshade setgray fill grestore 
               linewidth setlinewidth stroke
              }bind def
}%
\special{rawpostscript
/sharpcornersnt{%
               sharpboxnt gsave grayshade setgray fill grestore 
               linewidth setlinewidth stroke
              }bind def
}%
\special{rawpostscript
/sharpcornersnb{%
               sharpboxnb gsave grayshade setgray fill grestore 
               linewidth setlinewidth stroke
              }bind def
}%
\special{rawpostscript
/roundcorners{%
               roundbox gsave grayshade setgray fill grestore 
               linewidth setlinewidth stroke
              }bind def
}%
\special{rawpostscript
/plainbox{%
           sharpbox grayshade setgray fill 
          }bind def
}%
%
\special{rawpostscript
/roundnoframe{%
               roundbox grayshade setgray fill 
              }bind def
}%
\special{rawpostscript
/sharpnoframe{%
               sharpbox grayshade setgray fill 
              }bind def
}%
}%
%
%

\def\pshade#1{\Parashade{sharpcorners}{#1}{0.95}{10}{0.5}{10}{10}}


\def\boxit#1{\vbox{\hrule\hbox{\vrule\kern3pt
                                \vbox{\kern3pt#1\kern3pt}\kern3pt\vrule}\hrule}}

\def\boxitnb#1{\vbox{\hrule\hbox{\vrule\kern3pt
                                \vbox{\kern3pt#1\kern3pt}\kern3pt\vrule}}}

\def\boxitnt#1{\vbox{\hbox{\vrule\kern3pt
                                \vbox{\kern3pt#1\kern3pt}\kern3pt\vrule}\hrule}}

\long\def\boxtext#1#2{$$\boxit{\vbox{\hsize #1\noindent\strut #2\strut}}$$}
\long\def\boxtextnb#1#2{$$\boxitnb{\vbox{\hsize #1\noindent\strut #2\strut}}$$}
\long\def\boxtextnt#1#2{$$\boxitnt{\vbox{\hsize #1\noindent\strut #2\strut}}$$}

\def\texshopbox#1{\boxtext{462pt}{\vskip-1.5pc\pshade{\vskip-1.0pc#1\vskip-2.0pc}}}
\def\texshopboxnt#1{\boxtextnt{462pt}{\vskip-1.5pc\pshade{\vskip-1.0pc#1\vskip-2.0pc}}}
\def\texshopboxnb#1{\boxtextnb{462pt}{\vskip-1.5pc\pshade{\vskip-1.0pc#1\vskip-2.0pc}}}


%
%
%
%
%
%
%
%
\font\helbigbig=cmr10 scaled 2500%
\font\helbigb=cmbx10 scaled 1500%
\font\eightbold=cmbx8%

\def\tenf{\hel}%
\def\tenit{\heli}%
\def\ninef{\ninehel}%
\def\nineit{\nineheli}%
%
%


\font\tenrm=cmr10%
\font\teni=cmmi10%
\font\tensy=cmsy10%
\font\tenbf=cmbx10%
\font\tentt=cmtt10%
\font\tenit=cmti10%
\font\tensl=cmsl10%

\def\tenpoint{\def\rm{\fam0\tenrm}%
\textfont0=\tenrm%
\textfont1=\teni%
\textfont2=\tensy%
\textfont\itfam=\tenit%
\textfont\slfam=\tensl%
\textfont\ttfam=\tentt%
\textfont\bffam=\tenbf%
\scriptfont0=\sevenrm%
\scriptfont1=\seveni%
\scriptfont2=\sevensy%
\scriptscriptfont0=\sixrm%
\scriptscriptfont1=\sixi%
\scriptscriptfont2=\sixsy%
\def\it{\fam\itfam\tenit}%
\def\tt{\fam\ttfam\tentt}%
\def\sl{\fam\slfam\tensl}%
\scriptfont\bffam=\sevenbf%
\scriptscriptfont\bffam=\sixbf%
\def\bf{\fam\bffam\tenbf}%
\normalbaselineskip=18pt%
\normalbaselines\rm}%

\font\ninerm=cmr9%
\font\ninebf=cmbx9%
\font\nineit=cmti9%
\font\ninesy=cmsy9%
\font\ninei=cmmi9%
\font\ninett=cmtt9%
\font\ninesl=cmsl9%

\def\ninepoint{\def\rm{\fam0\ninerm}%
\textfont0=\ninerm%
\textfont1=\ninei%
\textfont2=\ninesy%
\textfont\itfam=\nineit%
\textfont\slfam=\ninesl%
\textfont\ttfam=\ninett%
\textfont\bffam=\ninebf%
\scriptfont0=\sixrm%
\scriptfont1=\sixi%
\scriptfont2=\sixsy%
\def\it{\fam\itfam\nineit}%
\def\tt{\fam\ttfam\ninett}%
\def\sl{\fam\slfam\ninesl}%
\scriptfont\bffam=\sixbf%
\scriptscriptfont\bffam=\fivebf%
\def\bf{\fam\bffam\ninebf}%
\normalbaselineskip=16pt%
\normalbaselines\rm}%

\font\eightrm=cmr8%
\font\eighti=cmmi8%
\font\eightsy=cmsy8%
\font\eightbf=cmbx8%
\font\eighttt=cmtt8%
\font\eightit=cmti8%
\font\eightsl=cmsl8%

\def\eightpoint{\def\rm{\fam0\eightrm}%
\textfont0=\eightrm%
\textfont1=\eighti%
\textfont2=\eightsy%
\textfont\itfam=\eightit%
\textfont\slfam=\eightsl%
\textfont\ttfam=\eighttt%
\textfont\bffam=\eightbf%
\scriptfont0=\sixrm%
\scriptfont1=\sixi%
\scriptfont2=\sixsy%
\scriptscriptfont0=\fiverm%
\scriptscriptfont1=\fivei%
\scriptscriptfont2=\fivesy%
\def\it{\fam\itfam\eightit}%
\def\tt{\fam\ttfam\eighttt}%
\def\sl{\fam\slfam\eightsl}%
\scriptscriptfont\bffam=\fivebf%
\def\bf{\fam\bffam\eightbf}%
\normalbaselineskip=14pt%
\normalbaselines\rm}%

\font\sevenrm=cmr7%
\font\seveni=cmmi7%
\font\sevensy=cmsy7%
\font\sevenbf=cmbx7%

\def\sevenpoint{%
   \def\rm{\sevenrm}\def\bf{\sevenbf}%
   \def\smc{\sevensmc}\baselineskip=12pt\rm}%

\font\sixrm=cmr6%
\font\sixi=cmmi6%
\font\sixsy=cmsy6%
\font\sixbf=cmbx6%

\fontdimen13\tensy=2.6pt%
\fontdimen14\tensy=2.6pt%
\fontdimen15\tensy=2.6pt%
\fontdimen16\tensy=1.2pt%
\fontdimen17\tensy=1.2pt%
\fontdimen18\tensy=1.2pt%

\def\tenf{\tenpoint}%
\def\ninef{\ninepoint}%
%



\def\tenpoint{\def\rm{\fam0\tenrm}%
\textfont0=\tenrm%
\textfont1=\teni%
\textfont2=\tensy%
\textfont\itfam=\tenit%
\textfont\slfam=\tensl%
\textfont\ttfam=\tentt%
\textfont\bffam=\tenbf%
\scriptfont0=\sevenrm%
\scriptfont1=\seveni%
\scriptfont2=\sevensy%
\scriptscriptfont0=\sixrm%
\scriptscriptfont1=\sixi%
\scriptscriptfont2=\sixsy%
\def\it{\fam\itfam\tenit}%
\def\tt{\fam\ttfam\tentt}%
\def\sl{\fam\slfam\tensl}%
\scriptfont\bffam=\sevenbf%
\scriptscriptfont\bffam=\sixbf%
\def\bf{\fam\bffam\tenbf}%
\normalbaselineskip=14pt%
\normalbaselines\rm}%

\def\ninepoint{\def\rm{\fam0\ninerm}%
\textfont0=\ninerm%
\textfont1=\ninei%
\textfont2=\ninesy%
\textfont\itfam=\nineit%
\textfont\slfam=\ninesl%
\textfont\ttfam=\ninett%
\textfont\bffam=\ninebf%
\scriptfont0=\sixrm%
\scriptfont1=\sixi%
\scriptfont2=\sixsy%
\def\it{\fam\itfam\nineit}%
\def\tt{\fam\ttfam\ninett}%
\def\sl{\fam\slfam\ninesl}%
\scriptfont\bffam=\sixbf%
\scriptscriptfont\bffam=\fivebf%
\def\bf{\fam\bffam\ninebf}%
\normalbaselineskip=13pt%
\normalbaselines\rm}%

\def\eightpoint{\def\rm{\fam0\eightrm}%
\textfont0=\eightrm%
\textfont1=\eighti%
\textfont2=\eightsy%
\textfont\itfam=\eightit%
\textfont\slfam=\eightsl%
\textfont\ttfam=\eighttt%
\textfont\bffam=\eightbf%
\scriptfont0=\sixrm%
\scriptfont1=\sixi%
\scriptfont2=\sixsy%
\scriptscriptfont0=\fiverm%
\scriptscriptfont1=\fivei%
\scriptscriptfont2=\fivesy%
\def\it{\fam\itfam\eightit}%
\def\tt{\fam\ttfam\eighttt}%
\def\sl{\fam\slfam\eightsl}%
\scriptscriptfont\bffam=\fivebf%
\def\bf{\fam\bffam\eightbf}%
\normalbaselineskip=12pt%
\normalbaselines\rm}%

\def\sevenpoint{%
   \def\rm{\sevenrm}\def\bf{\sevenbf}%
   \def\smc{\sevensmc}\baselineskip=10pt\rm}%

\def\section#1{\goodbreak\vskip 3pc plus 6pt minus 3pt\leftskip=-2pc
   \global\advance\sectnum by 1\eqnumber=1\subsectnum=0%
\global\examplnumber=1\figrnumber=1\propnumber=1\defnumber=1\lemmanumber=1\assumptionnumber=1 \conditionnumber =1%
   \line{\hfuzz=1pc{\hbox to 3pc{\bf 
   \vtop{\hfuzz=1pc\hsize=38pc\hyphenpenalty=10000\noindent\uppercase{\the\sectnum.\quad #1}}\hss}}
			\hfill}
			\leftskip=0pc\nobreak\tenf
			\vskip 1pc plus 4pt minus 2pt\noindent\ignorespaces}
\def\subsection#1{\noindent\leftskip=0pc\tenf
   \goodbreak\vskip 1pc plus 4pt minus 2pt
               \global\advance\subsectnum by 1
   \line{\hfuzz=1pc{\hbox to 3pc{\bf \the\sectnum.\the\subsectnum.   
   \vtop{\hfuzz=1pc\hsize=38pc\hyphenpenalty=10000\noindent{\bf #1}}\hss}}
                        \hfill}
   \leftskip=0pc\nobreak\tenf
                        \vskip 1pc plus 4pt minus 2pt\nobreak\noindent\ignorespaces}



\def\texshopbox#1{\boxtext{462pt}{\vskip-1.5pc\pshade{\vskip-1.0pc#1\vskip-2.0pc}}}
\def\texshopboxnt#1{\boxtextnt{462pt}{\vskip-1.5pc\pshade{\vskip-1.5pc#1\vskip-2.0pc}}}
\def\texshopboxnb#1{\boxtextnb{462pt}{\vskip-1.5pc\pshade{\vskip-1.0pc#1\vskip-2.5pc}}}


\input miniltx

\ifx\pdfoutput\undefined
  \def\Gin@driver{dvips.def} 
\else
  \def\Gin@driver{pdftex.def} 
\fi

\input graphicx.sty
\resetatcatcode

\long\def\fig#1#2#3{\vbox{\vskip1pc\vskip#1
\prevdepth=12pt \baselineskip=12pt
\vskip1pc
\hbox to\hsize{\hfill\vtop{\hsize=30pc\noindent{\eightbf Figure #2\ }
{\eightpoint#3}}\hfill}}}

\def\show#1{}

\rightheadline{\botmark}

\pageno=1

\rightheadline{\botmark}

\pn {\bf February 2014 (Revised August 2014, and June 2016)}\hfill{\bf Report LIDS - 2915}
\bigskip \bigskip\bigskip

\bigskip

\def\longpapertitle#1#2#3{{\bf \centerline{\helbigb
{#1}}}\medskip{\bf \centerline{\helbigb
{#2}}}\bigskip{\bf \centerline{
{#3}}}\bigskip}

\vskip-3pc

\def\jstar{J^{\raise0.04pt\hbox{\sevenpoint *}} }
\def\qstar{Q^{\raise0.04pt\hbox{\sevenpoint *}} }

\longpapertitle{Robust Shortest Path Planning and}{Semicontractive Dynamic Programming}{ {Dimitri P.\ Bertsekas\footnote{\dag}{\ninepoint  Dimitri Bertsekas is with the Dept.\ of Electr.\ Engineering and
Comp.\ Science, and the Laboratory for Information and Decision Systems, M.I.T., Cambridge, Mass., 02139.} }}

\centerline{\bf Abstract}

In this paper we consider shortest path problems in a directed graph where the transitions between nodes are subject to uncertainty. We use a minimax formulation, where the objective is to guarantee that a special destination state is reached with a minimum cost path under the worst possible instance of the uncertainty. Problems of this type arise, among others, in planning and pursuit-evasion contexts, and in model predictive control. Our analysis makes use of the recently developed theory of abstract semicontractive dynamic programming models. We investigate questions of existence and uniqueness of solution of the optimality equation, existence of optimal paths, and the validity of various algorithms patterned after the classical methods of value and policy iteration, as well as a Dijkstra-like algorithm for problems with nonnegative arc lengths.

\vskip-1pc

\vfill\eject

\section{Introduction}

\vskip-0.5pc

\pn In this paper, we discuss shortest path problems that embody a worst-case view of uncertainty. These problems relate to several other types of problems arising in stochastic and minimax control, model predictive control, Markovian decision processes, planning, sequential games, robust and combinatorial optimization, and solution of discretized large-scale differential equations. Consequently, our analysis and algorithms relate to a large body of existing theory. However, in this paper we rely on a recently developed abstract dynamic programming theory of semicontractive problems, and capitalize on general results developed in the context of this theory [Ber13]. We first discuss informally these connections and we survey the related literature.

\subsubsection{Relations with Other Problems and Literature Review}

\pn The closest connection to our work is the classical shortest path problem where the objective is to reach a destination node with a minimum length path from every other node in a directed graph. This is a fundamental problem that has an enormous range of applications and has been studied extensively (see e.g., the surveys [Dre69], [GaP88], and many textbooks, including [Roc84], [AMO89], [Ber98], [Ber05]). The assumption is that at any node $x$, we may determine a successor node $y$ from a given set of possible successors, defined by the arcs $(x,y)$ of the graph that are outgoing from $x$. 

In some problems, however, following the decision at a given node, there is inherent uncertainty about the successor node. In a stochastic formulation, the uncertainty is modeled by a probability distribution over the set of successors, and our decision is then to choose at each node one distribution out of a given set of distributions. 
The resulting problem, known as stochastic shortest path problem (also known as transient programming problem), is a total cost infinite horizon Markovian decision problem, with a substantial analytical and algorithmic methodology, which finds extensive applications in problems of motion planning, robotics, and other problems where the aim is to reach a goal state with probability 1 under stochastic uncertainty (see [Pal67], [Der70], [Pli78], [Whi82], [BeT89], [BeT91], [Put94], [HCP99], [HiW05], [JaC06], [LaV06], [Bon07], [Ber12], [BeY16], [CCV14], [YuB13a]). Another important area of application is large-scale computation for discretized versions of  differential equations (such as the Hamilton-Jacobi-Bellman equation, and the eikonal equation); see [GoR85], [Fal87], [KuD92], [BGM95], [Tsi95], [PBT98], [Set99a], [Set99b], [Vla08], [AlM12], [ChV12], [AnV14], [Mir14].

In this paper, we introduce a sequential minimax formulation of the shortest path problem, whereby the uncertainty is modeled by set membership: at a given node, we may choose one subset out of a given collection of subsets of nodes, and the successor node on the path is chosen from this subset by an antagonistic opponent. Our principal method of analysis is dynamic programming (DP for short).
Related problems have been studied for a long time, in the context of control of uncertain discrete-time dynamic systems with a set membership description of the uncertainty (starting with the theses [Wit66] and [Ber71], and followed up by many other works; see e.g., the monographs [BaB91], [KuV97], [BlM08], the survey [Bla99], and the references given there). These problems are relevant for example in the context of model predictive control under uncertainty, a subject of great importance in the current practice of control theory (see e.g., the surveys [MoL99], 
[MRR00], and the books [Mac02], [CaB04], [RaM09]; model predictive control with set membership disturbances is discussed in the thesis [Ker00] and the text [Ber05], Section 6.5.2). 

Sequential minimax problems have also been studied in the context of sequential games (starting with the paper [Sha53], and followed up by many other works, including the books [BaB91], [FiV96], and the references given there). Sequential games that involve shortest paths are particularly relevant; see the works [PaB99], [GrJ08], [Yu11], [BaL15].  An important difference with some of the works on sequential games is that in our minimax formulation, we assume that the antagonistic opponent knows the decision and corresponding subset of successor nodes chosen at each node. Thus in our problem, it would make a difference if the decisions at each node were made with advance knowledge of the opponent's choice (``min-max" is typically not equal to ``max-min" in our context). 
Generally shortest path games admit a simpler analysis when the arc lengths are assumed nonnegative (as is done for example in the recent works [GrJ08], [BaL15]), when the problem inherits the structure of negative DP (see [Str66], or the texts [Put94], [Ber12]) or abstract monotone increasing abstract DP models (see [Ber77], [BeS78], [Ber13]). However, our formulation and line of analysis is based on the recently introduced abstract semicontractive DP model of [Ber13], and allows negative as well as nonnegative arc lengths. Problems with negative arc lengths arise in applications when we want to find the longest path in a network with nonnegative arc lengths, such as critical path analysis. Problems with both positive and negative arc lengths include searching a network for objects of value with positive search costs (cf.\ Example 4.3), and financial problems of maximization of total reward when there are transaction and other costs.

An important application of our shortest path problems is in pursuit-evasion (or search and rescue) contexts, whereby a team of ``pursuers" are aiming to reach one or more ``evaders" that move unpredictably. Problems of this kind have been studied extensively from different points of view (see e.g., [Par76], [MHG88], [BaS91], [HKR93], [BSF94], [BBF95], [GLL99], [VKS02], [LaV06], [AHK07], [BBH08], [BaL15]). For our shortest path formulation to be applicable to such a problem, the pursuers and the evaders must have perfect information about each others' positions, and the Cartesian product of their positions (the state of the system) must be restricted to the finite set of nodes of a given graph, with known transition costs (i.e., a ``terrain map" that is known a priori). 

We may deal with pursuit-evasion problems with imperfect state information and set-membership uncertainty by means of a reduction to perfect state information, which is based on set membership estimators and the notion of a sufficiently informative function, introduced in the thesis [Ber71] and in the subsequent paper [BeR73]. In this context, the original imperfect state information problem is reformulated as a problem of perfect state information, where the states correspond to subsets of nodes of the original graph (the set of states that are consistent with the observation history of the system, in the terminology of set membership estimation [BeR71], [Ber71], [KuV97]). Thus, since $X$ has a finite number of nodes, the reformulated problem still involves a finite (but much larger) number of states, and may be dealt with using the methodology of this paper. Note that the problem reformulation just described is also applicable to general minimax control problems with imperfect state information, not just to pursuit-evasion problems.

Our work is also related to the subject of robust optimization (see e.g., the book [BGN09] and the recent survey [BBC11]), which includes minimax formulations of general optimization problems with set membership uncertainty. However, our emphasis here is placed on the presence of the destination node and the requirement for termination, which is the salient feature and the essential structure of shortest path problems. Moreover, a difference with other works on robust shortest path selection  (see e.g., [YuY98], [BeS03], [MoG04]) is that in our work the uncertainty about transitions or arc cost data at a given node is decoupled from the corresponding uncertainty at other nodes. This allows a DP formulation of our problem. 

Because our context differs in essential respects from the preceding works, the results of the present paper are new to a great extent. 
The line of analysis is also new, and is based on the connection with the theory of abstract semicontractive DP mentioned earlier. In addition to simpler proofs, a major benefit of this abstract line of treatment is deeper insight into the structure of our problem, and the nature of our analytical and computational results. Several related problems, involving for example an additional stochastic type of uncertainty, admit a similar treatment. Some of these problems are described in the last section, and their analysis and associated algorithms are subjects for further research.

\subsubsection{Robust Shortest Path Problem Formulation}

\pn
To formally describe our problem, we consider a graph with a finite set of nodes $X \cup\{t\}$ and a finite set of directed arcs ${\cal A}\subset\big\{(x,y)\mid x,y\in X \cup\{t\}\big\}$, where $t$ is a special node called the {\it destination\/}. At each node $x\in X$ we may choose a control or action $u$ from a nonempty set $U(x)$, which is  a subset of a finite set $U$. Then a successor node $y$ is selected by an antagonistic opponent from a nonempty set $Y(x,u)\subset X \cup\{t\}$, such that $(x,y)\in {\cal A}$ for all $y\in Y(x,u)$,  and a cost  $g(x,u,y)$ is incurred. The destination node $t$ is absorbing and cost-free, in the sense that the only outgoing arc from $t$ is $(t,t)$ and we have $g(t,u,t)=0$ for all $u\in U(t)$. 
 
A policy is defined to be a function $\m$ that assigns to each node $x\in X$ a control $\m(x)\in U(x)$. We denote the finite set of all policies by ${\cal M}$. A {\it possible  path under a policy $\m$ starting at  node $x_0\in X$} is an arc sequence of the form
$$p=\big\{(x_0,x_1),(x_1,x_2),\ldots\big\},$$
such that $x_{k+1}\in Y\big(x_k,\m(x_k)\big)$ for all $k\ge0$. The set of all possible  paths under $\m$ starting at $x_0$ is denoted by $P(x_0,\m)$; it is the set of paths that the antagonistic opponent may generate starting from $x$,  once policy $\m$ has been chosen. 
The length of a path $p\in P(x_0,\m)$ is defined by
$$L_{\m}(p)=\sum_{k=0}^{\infty} g\big(x_k, \m(x_k),x_{k+1}\big),$$
if the series above is convergent, and more generally by
$$L_{\m}(p)=\limsup_{m\to\infty}\sum_{k=0}^{m} g\big(x_k, \m(x_k),x_{k+1}\big),$$
if it is not. 
For completeness, we also define the length of a portion 
$$\big\{(x_i,x_{i+1}),(x_{i+1},x_{i+2}),\ldots,(x_m, x_{m+1})\big\}$$
 of a path $p\in P(x_0,\m)$, consisting of a finite number of consecutive arcs, by
$$\sum_{k=i}^{m}g\big(x_k, \m(x_k),x_{k+1}\big).$$
When confusion cannot arise we will also refer to such a finite-arc portion as a path.
Of special interest are {\it cycles\/}, that is, paths of the form $\big\{(x_i,x_{i+1}),(x_{i+1},x_{i+2}),\ldots,(x_{i+m},x_i)\big\}$. Paths that do not contain any cycle other than the self-cycle $(t,t)$ are called {\it simple\/}. 

For a given policy $\m$ and $x_0\ne t$, a path $p\in P(x_0,\m)$ is said to be {\it terminating} if it has the form
$$p=\big\{(x_0,x_1),(x_1,x_2),\ldots,(x_m,t),(t,t),\ldots\big\},\xdef\termpath{\lab}\eqnum\show{oneo}$$
where $m$ is a positive integer, and $x_0,\ldots,x_m$ are distinct nondestination nodes. Since $g(t,u,t)=0$ for all $u\in U(t)$, the  length of a terminating path $p$ of the form \termpath, corresponding to $\m$, is given by
$$L_{\m}(p)=g\big(x_m,\m(x_m),t\big)+\sum_{k=0}^{m-1} g\big(x_k, \m(x_k),x_{k+1}\big),$$
and is equal to the finite length of its initial portion that consists of the first $m+1$ arcs. 

\xdef \figsimplrobustt{\figr}\figrnum\show{myfigure}

An important characterization of a policy $\m$ is provided by the subset of arcs 
$${\cal A}_\m=\cup_{x\in X}\big\{(x,y)\mid y\in Y\big(x,\m(x)\big)\big\}.$$ 
Thus ${\cal A}_\m$, together with the self-arc $(t,t)$, consists of the set of paths $\cup_{x\in X}P(x,\m)$, in the sense that it contains this set of paths and no other paths. We say that ${\cal A}_\m$ is {\it destination-connected} if for each $x\in X$ there exists a terminating path in $P(x,\m)$. 
We say that $\m$ is {\it proper} if  the subgraph of arcs ${\cal A}_\m$ 
is acyclic (i.e., contains no cycles). Thus $\m$ is proper if and only if all the paths in $\cup_{x\in X}P(x,\m)$ are simple and hence terminating (equivalently $\m$ is proper if and only if ${\cal A}_\m$ is destination-connected and has no cycles). The term ``proper" is consistent with a similar term in stochastic shortest path problems, where it indicates a policy under which the destination is reached with probability 1, see e.g., [Pal67], [BeT89], [BeT91]. If $\m$ is not proper, it is called {\it improper\/}, in which case the subgraph of arcs ${\cal A}_\m$ must contain a cycle; see the examples of Fig.\ \figsimplrobustt. 

\topinsert
\centerline{\hskip0pc\includegraphics[width=4.2in]{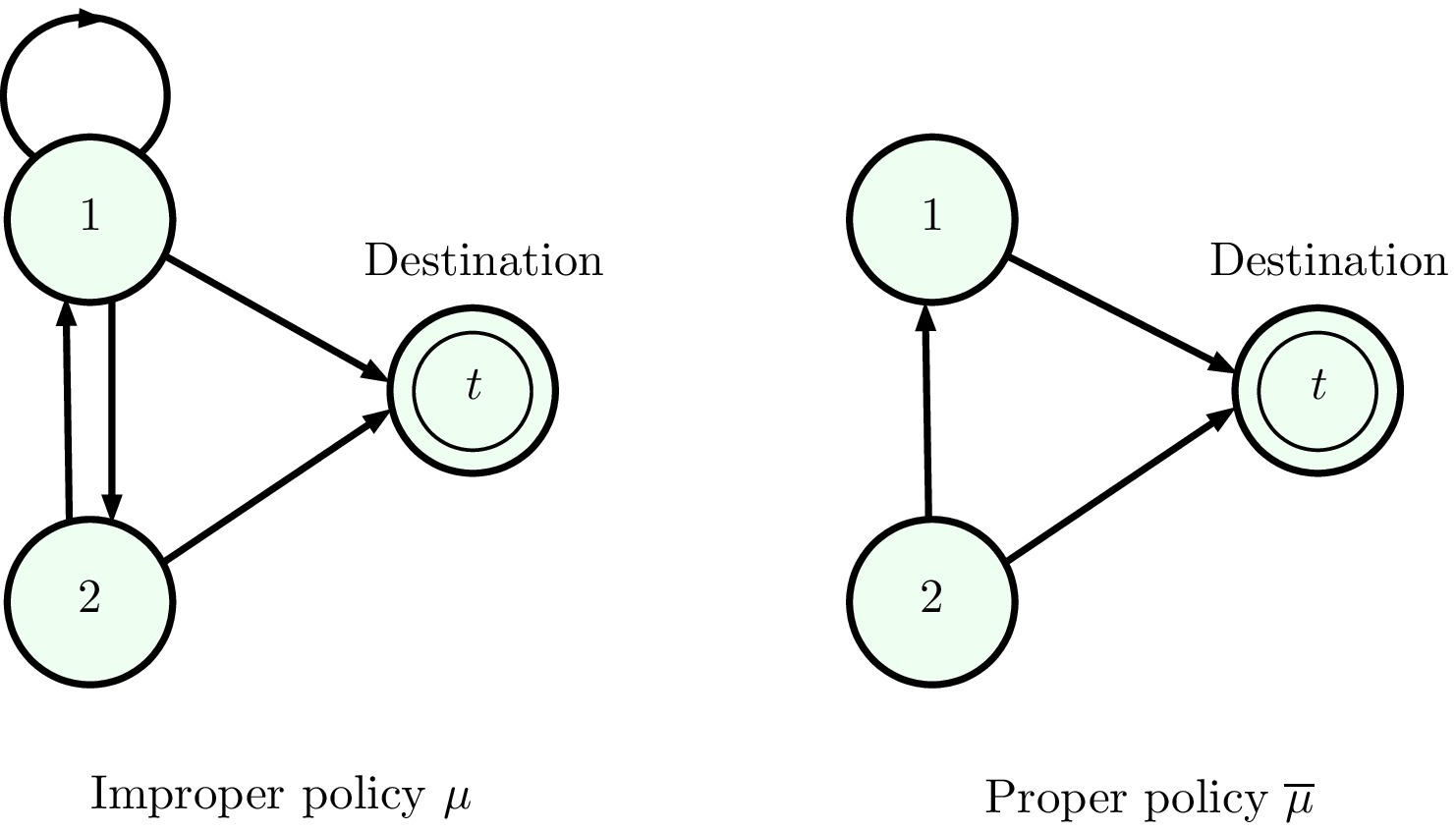}}
\vskip-0.5pc
\hskip-4pc\fig{0pc}{\figsimplrobustt.} {A robust shortest path problem with $X=\{1,2\}$, two controls at node 1, and one control at node 2. There are two policies, $\m$ and $\ol\m$, corresponding to the two controls at node 1. The figure shows the subgraphs of arcs ${\cal A}_\m$ and ${\cal A}_{\ol\m}$. The policy $\m$ is improper because ${\cal A}_\m$ contains the cycle $(1,2,1)$ and the (self-)cycle $(1,1)$.}
\endinsert


For a proper $\m$, we associate with every $x\in X$ the worst-case path length over the finite set of possible paths starting from $x$, which is denoted by
$$J_\m(x)=\max_{p\in P(x,\m)}L_{\m}(p),\qquad x\in X.\xdef\propercost{\lab}\eqnum\show{oneo}$$ 
Thus $J_\m(x)$ is the length of the {\it longest} path from $x$ to $t$ in the acyclic subgraph of arcs ${\cal A}_\m$. Since there are finitely many paths in this acyclic graph, $J_\m(x)$ may be found either by enumeration and comparison of these paths (in simple cases), or by solving the shortest path problem obtained when the signs of the arc lengths $g\big(x,\m(x),y\big)$, $(x,y)\in {\cal A}_\m$, are reversed.

Our problem is to find an optimal proper policy, i.e., one that minimizes $J_\m(x)$ over all proper $\m$, simultaneously for all $x\in X$, under assumptions that parallel those for the classical shortest path problem. We refer to this as the problem of {\it robust shortest path} selection (RSP for short). Note that in our problem, {\it reaching the destination starting from every node is a requirement\/}, regardless of the choices of the hypothetical antagonistic opponent. In other words the minimization
 in RSP is over the proper policies only.
 
Of course for the problem to have a feasible solution and thus be meaningful, there must exist at least one proper policy, and this may be restrictive for a given problem.
One may deal with cases where feasibility is not known to hold by introducing  for every $x$ an artificial ``termination action" $\ol u$ into $U(x)$ [i.e., a $\ol u$ with $Y(x,\ol u)=\{t\}$], associated with very large length [i.e., $g(x,\ol u,t)=\ol g>>1$]. Then the policy $\ol\m$ that selects the termination action at each $x$ is proper and has cost function $J_{\ol\m}(x)\equiv\ol g$. In the problem thus reformulated the optimal cost over proper policies will be unaffected for all nodes $x$ for which there exists a proper policy $\m$ with $J_\m(x)<\ol g$. Since for a proper $\m$, the cost $J_\m(x)$ is bounded above by the number of nodes in $X$ times the largest arc length, a suitable value of $\ol g$ is readily available.

In Section 2 we will formulate RSP in a way that the semicontractive DP framework can be applied. In Section 3, we will describe briefly this framework and we will quote the results that will be useful to us. In Section 4, we will develop our main analytical results for RSP. In Section 5, we will discuss algorithms of the value and policy iteration type, by specializing corresponding algorithms of semicontractive DP, and by adapting available algorithms for stochastic shortest path problems. Among others, we will give a Dijkstra-like algorithm for problems with nonnegative arc lengths, which terminates in a number of iterations equal to the number of nodes in the graph, and has low order polynomial complexity. Related Dijkstra-like algorithms were proposed recently, in the context of dynamic games and with an abbreviated convergence analysis, by [GrJ08] and [BaL15]. 

\vskip-1pc
\section{Minimax Formulation}
\vskip-0.5pc

\pn In this section we will reformulate RSP into a minimax problem, whereby given a policy $\m$, an antagonistic opponent selects a successor node $y\in Y\big(x,\m(x)\big)$ for each $x\in X$, with the aim of maximizing the lengths of the resulting paths. The essential difference between RSP and the associated minimax problem is that {\it in RSP only the proper policies are admissible, while in the minimax problem all policies will be admissible\/}. Our analysis will be based in part on assumptions under which improper policies cannot be optimal for the minimax problem, implying that optimal policies for the minimax problem will be optimal for the original RSP problem. 
One such assumption is the following.

\xdef\assumptionone{\assumptionn}\assumptionnum\show{myproposition}

\texshopbox{\assumption{\assumptionone:}
\nitem{(a)} There exists at least one proper policy.
\nitem{(b)} For every improper policy $\m$, all cycles in the subgraph of arcs ${\cal A}_\m$ have positive length.
}

The preceding assumption parallels and generalizes the typical assumptions in the classical deterministic shortest path problem, i.e., the case where $Y(x,\m)$ consists of a single node. Then condition (a) is equivalent to assuming that each node is connected to the destination with a path, while condition (b) is equivalent to assuming that all directed cycles in the graph have positive length.\footnote{\dag}{\ninepoint To verify the existence of a proper policy [condition (a)] one may apply a reachability algorithm, which  constructs the sequence $\{N_k\}$ of  sets 
$$N_{k+1}=N_k\cup \big\{x\in X\cup\{t\}\mid \hbox{there exists $u\in U(x)$ with }Y(x,u)\subset N_k\big\},$$
starting with $N_0=\{t\}$ (see [Ber71], [BeR71]). A proper policy exists if and only if this algorithm stops with a final set $\cup_k N_k$ equal to $X\cup\{t\}$. If there is no proper policy, this algorithm will stop with $\cup_k N_k$ equal to a strict subset of $X\cup\{t\}$ of nodes starting from which there exists a terminating path under some policy. The problem may then be reformulated over the reduced graph consisting of the node set $\cup_k N_k$, so there will exist a proper policy in this reduced problem.}
 Later in Section 4, in addition to Assumption \assumptionone, we will consider another weaker assumption, whereby ``positive length" is replaced with ``nonnegative length" in condition (b) above. This assumption will hold in the common case where all arc lengths $g(x,u,y)$ are nonnegative, but there may exist a zero length cycle.
As a first step, we extend the definition of the function $J_\m$ to the case of an improper policy. 
Recall that for a proper policy $\m$, $J_\m(x)$ has been defined by Eq.\ \propercost, as the  length of the longest path $p\in P(x,\m)$,
$$J_\m(x)=\max_{p\in P(x,\m)}L_{\m}(p),\qquad x\in X.\xdef\propercosta{\lab}\eqnum\show{oneo}$$ 
We extend this definition to any policy $\m$, proper or improper, by defining $J_\m(x)$ as
$$J_\m(x)=\limsup_{k\to \infty}\sup_{p\in P(x,\m)}L_p^k(\m),\xdef\jmimprop{\lab}\eqnum\show{oneo}$$
where $L_p^k(\m)$ is the sum of lengths of the first $k$ arcs in the path $p$. 
When $\m$ is proper, this definition coincides with the one given earlier [cf.\ Eq.\ \propercosta]. Thus for a proper $\m$, $J_\m$ is real-valued, and it is the unique solution of the optimality equation (or Bellman equation) for the longest path problem associated with the proper policy $\m$ and the acyclic subgraph of arcs ${\cal A}_\m$:
$$J_\m(x)=
\max_{y\in Y(x,\m(x))}\big[g\big(x,\m(x),y\big)+\tl J_\m(y)\big]\qquad x\in X,\xdef\belleq{\lab}\eqnum\show{oneo}$$
where we denote by $\tl J_\m$ the function given by
$$\tl J_\m(y)=\cases{J_\m(y)&if $y\in X$,\cr
0&if $y=t$.\cr}\xdef\tmmapdspotz{\lab}\eqnum\show{oneo}$$
Any shortest path algorithm may be used to solve this longest path problem for a proper $\m$.
However, when $\m$ is improper, we may have $J_\m(x)=\infty$, and the solution of the corresponding longest path problem may be problematic. 

We will  consider the problem of finding
$$\jstar(x)=\min_{\m\in {\cal M}}J_\m(x),\qquad x\in X,\xdef\optcost{\lab}\eqnum\show{oneo}$$
and a policy attaining the minimum above, simultaneously for all $x\in X$. Note that {\it the minimization is over all policies, in contrast with the RSP problem, where the minimization is over just the proper policies\/}.

\subsubsection{Embedding Within an Abstract DP Model}

\pn We will now reformulate the minimax problem of Eq.\ \optcost\ more abstractly, by expressing it in terms of the mapping that appears in Bellman's equation \belleq-\tmmapdspotz, thereby bringing to bear the theory of abstract DP. We denote by $E(X)$ the set of functions $J: X \mapsto[-\infty,\infty]$, and by $R(X)$ the set of functions $J: X \mapsto(-\infty,\infty)$. Note that since $X$ is finite, $R(X)$ can be viewed as a finite-dimensional Euclidean space. We introduce the mapping $H:X\times U\times E(X)\mapsto [-\infty,\infty]$ given by
$$H(x,u,J)=
\max_{y\in Y(x,u)}\big[g(x,u,y)+\tl J(y)\big],\xdef\tmmapdspo{\lab}\eqnum\show{oneo}$$
where for any $J\in E(X)$ we denote by $\tl J$ the function given by
$$\tl J(y)=\cases{J(y)&if $y\in X$,\cr
0&if $y=t$.\cr}\eqnum\show{oneo}$$
We consider  for each policy $\m$, the mapping $T_\m: E(X)\mapsto  E(X)$, defined by
$$(T_\m J)(x)=H\big(x,\m(x),J\big),\qquad x\in X,\xdef\tmmapdspt{\lab}\eqnum\show{oneo}$$
and we note that the fixed point equation $J_\m=T_\m J_\m$ is identical to the Bellman equation \belleq.
We also consider the mapping $T:E(X)\mapsto  E(X)$ defined by
$$(TJ)(x)=\min_{u\in U(x)}H(x,u,J),\qquad x\in X,\xdef\tmapdspt{\lab}\eqnum\show{oneo}$$
also equivalently written as
$$(T J)(x)=\min_{\m\in{\cal M}}(T_\m J)(x),\qquad x\in X.\xdef\tmapdsptalt{\lab}\eqnum\show{oneo}$$
We denote by $T^k $ and $T_\m^k$ the $k$-fold compositions of the mappings $T$ and $T_\m $ with themselves, respectively. 

Let us consider the zero function, which we denote by $\bar J$:
$$\bar J(x)\equiv 0,\qquad x\in X.$$
Using Eqs.\ \tmmapdspo-\tmmapdspt, we see that for any $\m\in{\cal M}$ and $x\in X$,  $(T^k_\m\bar J)(x)$ is the result of the $k$-stage DP algorithm that computes $\sup_{p\in P(x,\m)}L_p^k(\m)$, the length of the longest path under $\m$ that starts at $x$ and consists of $k$ arcs, so that
$$(T_\m^k \bar J)(x)=\sup_{p\in P(x,\m)}L_p^k(\m),\qquad x\in X.$$
Thus the definition \jmimprop\ of $J_\m$ can be written in the alternative and equivalent form
$$J_\m(x)=\limsup_{k\to\infty}\,(T^k_\m\bar J)(x),\qquad x\in X.\xdef\jmimproper{\lab}\eqnum\show{oneo}$$
We are focusing on optimization over stationary policies because under the assumptions of this paper (both Assumption \assumptionone\ and the alternative assumptions of Section 4) the optimal cost function would not be improved by allowing nonstationary policies, as shown in [Ber13], Chapter 3.\footnote{\dag}{\ninepoint In the more general framework of [Ber13], nonstationary Markov policies of the form $\p=\{\m_0,\m_1,\ldots\}$, with $\m_k\in{\cal M}$, $k=0,1,\ldots$, are allowed, and their  cost function is defined by
$$J_\p(x)=\limsup_{k\to\infty}\,(T_{\m_0}\cdots T_{\m_{k-1}} \skew5\bar J)(x),\qquad x\in X,$$
where $T_{\m_0}\cdots T_{\m_{k-1}}$ is the composition of the mappings $T_{\m_0},\ldots,T_{\m_{k-1}}$. Moreover, $J^*(x)$ is defined as the infimum of $J_\p(x)$ over all such $\p$. However, under the assumptions of the present paper, this infimum is attained by a stationary policy (in fact one that is proper). Hence, attention may be restricted to stationary policies without loss of optimality and without affecting the results from [Ber13] that will be used.}

The results that we will show under Assumption \assumptionone\ generalize the main analytical results for the classical deterministic shortest path problem, and stated in abstract form, are the following:
\nitem{(a)} $\jstar$ is the unique fixed point of $T$ within $R(X)$, and we have $T^k J\to\jstar$ for all $J\in R(X)$.
\nitem{(b)} Only proper policies can be optimal, and there exists an optimal proper policy.\footnote{\dag}{\ninepoint  Since the set of policies is finite, there exists a policy minimizing $J_\m(x)$ over the set of proper policies $\m$, for each $x\in X$. However, the assertion here is stronger, namely that there exists a proper $\m^*$ minimizing $J_\m(x)$ over all $\m\in{\cal M}$ and {\it simultaneously for all $x\in X$\/}, i.e., a proper $\m^*$ with $J_{\m^*}=J^*$.}
\nitem{(c)} A policy $\m$ is optimal if and only if it attains the minimum for all $x\in X$ in Eq.\ \tmapdsptalt\ when $J=\jstar$.
\smskip
\pn Proofs of these results from first principles are quite complex. However, fairly easy proofs can be obtained by embedding the problem of minimizing the function $J_\m$ of Eq.\ \jmimproper\ over $\m\in{\cal M}$,  within the  abstract semicontractive DP framework introduced in [Ber13]. In particular, we will use general results for this framework, which we will summarize in the next section.

\vskip-1pc

\section{Semicontractive DP Analysis}

\pn We will now view the problem of minimizing over $\m\in {\cal M}$ the cost function $J_\m$,  given in the abstract form \jmimproper, as a special case of a semicontractive DP model.
We first provide a brief review of this model, with a notation that corresponds to the one used in the preceding section. 

The starting point is a set of states $X$, a set of controls $U$, and a control constraint set $U(x)\subset U$ for each $x\in X$. For the general framework of this section, $X$ and $U$ are arbitrary sets; we continue to use some of the notation of the preceding section in order to indicate the relevant associations. A policy is a mapping $\m:X\mapsto U$ with $\m(x)\in U(x)$ for all $x\in X$, and the set of all policies is denoted by ${\cal M}$. For each policy $\m$, we are given a mapping $T_\m:E(X)\mapsto E(X)$ that is monotone in the sense that for any two $J,J'\in E(X)$,
$$J\le J'\qquad \implies\qquad T_\m J\le T_\m J'.$$
We define the mapping $T:E(X)\mapsto E(X)$ by
$$(T J)(x)=\inf_{\m\in{\cal M}}(T_\m J)(x),\qquad x\in X.$$
The cost function of $\m$ is defined as
$$J_\m(x)=\limsup_{m\to\infty}\,(T_\m^k\bar J)(x),\qquad x\in X,$$
where $\bar J$ is some given function in $E(X)$. The objective is to find 
$$\jstar(x)=\inf_{\m\in{\cal M}}J_\m(x)$$
  for each $x\in X$, and a policy $\m$ such that $J_\m=\jstar$, if one exists. Based on the correspondences with Eqs.\ \tmmapdspo-\jmimproper, it can be seen that the minimax problem of the preceding section is the special case of the problem of this section, where $X$ and $U$ are finite sets, $T_\m$ is defined by Eq.\ \tmmapdspt, and $\bar J$ is the zero function. 

In contractive models, the mappings $T_\m$ are assumed to be contractions, with respect to a common weighted sup-norm and with a common contraction modulus, in the subspace of functions in $E(X)$ that are bounded with respect to the weighted sup-norm. These models have a strong analytical and algorithmic theory, which dates to [Den67]; see also [BeS78], Ch.\ 3, and recent extensive treatments given in Chapters 1-3 of [Ber12], and Ch.\ 2 of [Ber13]. In semicontractive models, only some policies have a contraction-like property. This property is captured by the notion of $S$-regularity of a policy introduced in [Ber13] and defined as follows. 

\texshopbox{
\pn {\bf Definition 3.1:} Given a set of functions $S\subset E(X)$, we say that a policy $\m$ is  {\it $S$-regular} if:
\nitem{(a)} $J_\m\in S$ and $J_\m=T_\m J_\m$.
\nitem{(b)} $\lim_{k\to\infty}T^k_\m J= J_\m$ for all $J\in S$. 
\smskip
\pn A policy that is not $S$-regular is called {\it $S$-irregular\/}.
}
\smskip

 Roughly, $\m$ is $S$-regular if $J_\m$ is an asymptotically stable equilibrium point of $T_\m$ within $S$. An important case of an $S$-regular  $\m$ is when $S$ is a complete subset of a metric space and $T_\m$ maps $S$ to $S$ and, when restricted to $S$, is a contraction with respect to the metric of that space.

There are several different choices of $S$, which may be useful depending on the context, such as for example $R(X)$, $E(X)$, $\big\{J\in R(X)\mid J\ge\bar J\big\}$, $\big\{J\in E(X)\mid J\ge\bar J\big\}$, and others. There are also several sets of assumptions and corresponding results, which are given in [Ber13] and will be used to prove our analytical results for the RSP problem. In this paper, we will use $S=R(X)$, but for ease of reference, we will quote results from [Ber13] with $S$ being an arbitrary subset of $R(X)$. 

We give below an assumption relating to semicontractive models, which is Assumption 3.2.1 of [Ber13]. A key part of this assumption is part (c), which implies that $S$-irregular policies have infinite cost for at least one state $x$, so they cannot be optimal. This part will provide a connection to Assumption \assumptionone(b).

\xdef\assumptiontoz{\assumptionn}\assumptionnum\show{myproposition}

\texshopboxnb{
\assumption{\assumptiontoz:} In the semicontractive model of this section with a set $S\subset R(X)$ the following hold:
\nitem{(a)} $S$ contains $\bar J$, and has the property that if $J_1,J_2$ are two functions in $S$, then $S$ contains all functions $J$ with $J_1\le J\le J_2$.
\nitem{(b)} The function $\hat J$ given by
$$\hat J(x)=\inf_{\m:\, S\hbox{\ninepoint -regular}}J_\m(x),\qquad x\in X,$$
belongs to $S$.
\nitem{(c)} For each $S$-irregular policy $\m$ and each $J\in S$, there is at least one state $x\in X$ such that 
$$\limsup_{k\to\infty}\,(T_\m^kJ)(x)=\infty.$$}\texshopboxnt{\nitem{} 
\nitem{(d)} The control set $U$ is a metric space, and the set
$$
\big\{ \m(x)\mid (T_\m J)(x)\le
\l\big\}$$
is compact for every
$J\in S$, $x\in X$, and $\l\in \re$.
\nitem{(e)} For each sequence  $\{J_m\}\subset S$ with $J_m\uparrow J$ for some $J\in S$
we have
$$\lim_{m\to\infty}(T_\m J_m)(x)=(T_\m J)(x),\qquad \forall\ x\in X,\ \m\in {\cal M}.$$
\nitem{(f)} For each  function $J\in S$, there exists a function  $J'\in S$ such that $J'\le J$ and $J'\le TJ'$.
\smskip
}

The following two propositions are given in [Ber13] as Prop.\ 3.2.1 and Lemma 3.2.4, respectively.\footnote{\dag}{\ninepoint As noted in the preceding section, a more general problem is defined in [Ber13], whereby nonstationary Markov policies are allowed, and $J^*$ is defined as the infimum over these policies. However,  under our assumptions, attention may be restricted to stationary policies without loss of optimality and without affecting the validity of the two propositions.} 
Our analysis will be based on these two propositions. 

\xdef\propabstracto{\propn}\propnum\show{myproposition}

\texshopbox{\proposition{\propabstracto:} Let Assumption \assumptiontoz\ hold. Then:
\nitem{(a)} The optimal cost function $\jstar$ is the unique fixed point of $T$ within the set $S$.
\nitem{(b)} A policy  $\m^*$ is optimal if and only if $T_{\m^*} \jstar =T\jstar$. Moreover, there exists an optimal $S$-regular policy.
\nitem{(c)} We have $T^k J\to \jstar$ for all $J\in S$. 
\nitem{(d)} For any $J\in S$, if $J\le TJ$ we have $J\le \jstar $, and if $J\ge TJ$ we have $J\ge \jstar $.
}

\xdef\propabstractt{\propn}\propnum\show{myproposition}

\texshopbox{\proposition{\propabstractt:} Let Assumption \assumptiontoz(b),(c),(d) hold.
Then:
\nitem{(a)} The function $\hat J$ of Assumption \assumptiontoz(b) is the unique fixed point of $T$ within $S$.
\nitem{(b)} Every policy $\m$ satisfying $T_{\m}\hat J=T\hat J$ is optimal within the set of $S$-regular policies, i.e., $\m$ is $S$-regular and $J_{\m}=\hat J$. Moreover, there exists at least one such policy.
}

The second proposition is useful for situations where only some of the conditions of Assumption \assumptiontoz\ are satisfied, and will be useful in the proof of an important part of Prop.\ 4.3 in the next section.

\section{Semicontractive Models and Shortest Path Problems}

\pn We will now apply the preceding two propositions to the minimax formulation of the RSP problem: minimizing over all $\m\in{\cal M}$  the shortest path cost $J_\m(x)$ as given by Eq.\ \jmimprop\ for both proper and improper policies. We will first derive some preliminary results. The following proposition clarifies the properties of $J_\m$ when $\m$ is improper.

\xdef\proppropimprop{\propn}\propnum\show{myproposition}

\texshopbox{\proposition{\proppropimprop:} Let $\m$ be an improper policy and let $J_\m$ be its  cost function as given by Eq.\ \jmimprop.
\nitem{(a)} If all cycles in the subgraph of arcs ${\cal A}_\m$ have nonpositive length, $J_\m(x)<\infty$ for all $x\in X$.
\nitem{(b)} If all cycles in the subgraph of arcs ${\cal A}_\m$ have nonnegative length, $J_\m(x)>-\infty$ for all $x\in X$.
\nitem{(c)} If all cycles in the subgraph of arcs ${\cal A}_\m$ have zero length, $J_\m$ is real-valued.
\nitem{(d)} If there is a positive length cycle in the subgraph of arcs ${\cal A}_\m$, we have $J_\m(x)=\infty$ for at least one node $x\in X$. More generally, for each $J\in R(X)$, we have $\limsup_{k\to\infty}(T^k_\m J)(x)=\infty$ for at least one $x\in X$.
}

\proof Any path with a finite number of arcs, can be decomposed into a simple path, and a finite number of cycles (see e.g., the path decomposition theorem of [Ber98], Prop.\ 1.1, and Exercise 1.4). Since there is only a finite number of simple paths under $\m$, their length is bounded above and below. Thus in part (a) the length of all paths with a finite number of arcs is bounded above, and in part (b) it is bounded below, implying that $J_\m(x)<\infty$ for all $x\in X$ or $J_\m(x)>-\infty$ for all $x\in X$, respectively. Part (c) follows by combining parts (a) and (b).  

To show part (d), consider a path $p$, which consists of an infinite repetition of the positive length cycle that is assumed to exist. Let $C_\m^k(p)$ be the length of the path that consists of the first $k$ cycles in $p$. Then  $C_\m^k(p)\to \infty$ and $C_\m^k(p)\le J_\m(x)$ for all $k$ [cf.\ Eq.\ \jmimprop], where $x$ is the first node in the cycle, thus implying that $J_\m(x)=\infty$. Moreover for every $J\in R(X)$ and all $k$, $(T^k_\m J)(x)$ is the maximum over the lengths of the $k$-arc paths that start at $x$, plus a terminal cost that is equal to either $J(y)$ (if the terminal node of the $k$-arc path is $y\in X$), or 0  (if the terminal node of the $k$-arc path  is the destination). Thus we have,
$$(T^k_\m \bar J)(x)+\min\lf\{0,\,\min_{x\in X}J(x)\ri\}\le (T^k_\m J)(x).$$
Since $\limsup_{k\to\infty}(T^k_\m \bar J)(x)=J_\m(x)=\infty$ as shown earlier, it follows that $\limsup_{k\to\infty}(T^k_\m J)(x)=\infty$ for all $J\in R(X)$.
\qed

Note that if there is a negative length cycle in the subgraph of arcs ${\cal A}_\m$, it is not necessarily true that for some $x\in X$ we have $J_\m(x)=-\infty$. Even for $x$ on the negative length cycle, the value of $J_\m(x)$ is determined by the {\it longest} path in $P(x,\m)$, which may be simple in which case $J_\m(x)$ is a real number, or contain an infinite repetition of a positive length cycle in which case $J_\m(x)=\infty$.

A key fact in our analysis is the following characterization of the notion of $R(X)$-regularity and its connection to the notion of properness. It shows that proper policies are $R(X)$-regular, but the set of $R(X)$-regular policies may contain some improper policies, which are characterized in terms of the sign of the lengths of their associated cycles.

\xdef\propproperchar{\propn}\propnum\show{myproposition}

\texshopbox{\proposition{\propproperchar:} Consider the minimax formulation of the RSP problem,  viewed as a special case of the abstract semicontractive DP model of Section 3.1 with $T_\m$ given by Eqs.\ \tmmapdspo-\tmmapdspt, and $\bar J$ being the zero function. The following are equivalent for a policy $\m$:
\nitem{(i)} $\m$ is $R(X)$-regular.
\nitem{(ii)} The subgraph of arcs ${\cal A}_\m$ is destination-connected and all its cycles have negative length. 
\nitem{(iii)} $\m$ is either proper or else, if it is improper, all the cycles of the subgraph of arcs ${\cal A}_\m$ have negative length, and $J_\m\in R(X)$.
} 

\proof To show that (i) implies (ii), let $\m$ be $R(X)$-regular and to arrive at a contradiction, assume that ${\cal A}_\m$ contains a nonnegative length cycle. Let $x$ be a node on the cycle, consider the path $p$ that starts at $x$ and consists of an infinite repetition of this cycle, and let $L_\m^k(p)$ be the length of the first $k$ arcs of that path. Let also $J$ be a nonzero constant function, $J(x)\equiv r$, where $r$ is a scalar. Then we have
$$L_\m^k(p)+r\le (T^k_\m J)(x),$$
since from the definition of $T_\m$, we have that $(T^k_\m J)(x)$ is the maximum over the lengths of all $k$-arc paths under $\m$ starting at $x$, plus $r$, if the last node in the path is not the destination. Since $\m$ is $R(X)$-regular, we have
$\limsup_{k\to\infty} (T^k_\m J)(x)=J_\m(x)<\infty$, so that for all scalars $r$,
$$\limsup_{k\to\infty}\,\big(L_\m^k(p)+r\big)\le J_\m(x)<\infty.$$
Taking infimum over $r\in \re$,  it follows that $\limsup_{k\to\infty}L_\m^k(p)=-\infty$, which contradicts the nonnegativity of the cycle of $p$. Thus all cycles of ${\cal A}_\m$ have negative length. 
To show that ${\cal A}_\m$ is destination-connected, assume the contrary. Then there exists some node $x\in X$ such that all paths in $P(x,\m)$ contain an infinite number of cycles. Since the length of all cycles is negative, as just shown, it follows that $J_\m(x)=-\infty$, which contradicts the $R(X)$-regularity of $\m$.

To show that (ii) implies (iii), we assume that $\m$ is improper and show that $J_\m\in R(X)$. By (ii) ${\cal A}_\m$ is destination-connected, so the set $P(x,\m)$ contains a simple path for all $x\in X$. Moreover, since by (ii) the cycles  of ${\cal A}_\m$ have negative length, each path in $P(x,\m)$ that is not simple has smaller length than some simple path in $P(x,\m)$. This implies that $J_\m(x)$ is equal to the largest path length among simple paths in $P(x,\m)$, so $J_\m(x)$ is a real number for all $x\in X$. 

To show that (iii) implies (i), we note that if $\m$ is proper, it is $R(X)$-regular, so we focus on the case where $\m$ is improper. Then by (iii), $J_\m\in R(X)$, so to show $R(X)$-regularity of $\m$, 
we must show that $(T^k_\m J)(x)\to J_\m(x)$ for all $x\in X$ and $J\in R(X)$, and that $J_\m=T_\m J_\m$. Indeed, from the definition of $T_\m$, we have 
$$(T^k_\m J)(x)=\sup_{p\in P(x,\m)}\big[L_\m^k(p)+J(x_p^k)\big],\xdef\tkmuj{\lab}\eqnum\show{oneo}$$
where $x_p^k$ is the node reached after $k$ arcs along the path $p$, and $J(t)$ is defined to be equal to 0. Thus as $k\to\infty$, for every path $p$ that contains an infinite number of cycles (each necessarily having negative length), the sequence $L_p^k(\m)+J(x_p^k)$ approaches $-\infty$. It follows that for sufficiently large $k$, the supremum in Eq.\ \tkmuj\ is attained by one of the simple paths in $P(x,\m)$, so $x_p^k=t$ and $J(x_p^k)=0$. Thus the limit of $(T^k_\m J)(x)$ does not depend on $J$, and is equal to the limit of $(T^k_\m \bar J)(x)$, i.e., $J_\m(x)$. To show that $J_\m=T_\m J_\m$, we note that by the preceding argument, $J_\m(x)$ is the length of the longest path among paths that start at $x$ and terminate at $t$. Moreover, we have 
$$(T_\m J_\m)(x)=\max_{y\in Y(x,\m(x))}\big[g(x,\m(x),y)+J_\m(y)\big],$$
where we denote $J_\m(t)=0$. Thus $(T_\m J_\m)(x)$ is also the length of the longest path among paths that start at $x$ and terminate at $t$, and hence it is equal to $J_\m(x)$. 
 \qed

We illustrate the preceding proposition with a two-node example involving an improper policy with a cycle that may have positive, zero, or negative length. 

\xdef\exampleirrego{\exampl}\examplnum\show{myexample}

\xdef \figsimplrobusto{\figr}\figrnum\show{myfigure}

\beginexample{\exampleirrego:}Let $X=\{1\}$, and consider the policy $\m$ where at state 1, the antagonistic opponent may force either staying at 1 or terminating, i.e., $Y\big(1,\m(1)\big)=\{1,t\}$. Then $\m$ is improper since its subgraph of arcs ${\cal A}_\m$ contains the self-cycle $(1,1)$; cf.\ Fig.\ \figsimplrobusto.  Let 
$$g\big(1,\m(1),1\big)=a,\qquad g\big(1,\m(1),t\big)=0.$$
 Then, 
$$(T_\m J_\m)(1)= \max\,\big[0,\,a+J_\m(1)\big],$$
and 
$$J_\m(1)=\cases{\infty&if $a>0$,\cr
0&if $a\le0$.\cr}$$
Consistently with Prop.\ \propproperchar, the following hold:
\nitem{(a)} For $a>0$, the cycle $(1,1)$ has positive length, and $\m$ is $R(X)$-irregular because $J_\m(1)=\infty$.
\nitem{(b)} For $a=0$, the cycle $(1,1)$ has zero length, and $\m$ is $R(X)$-irregular because for a function $J\in R(X)$ with $J(1)>0$, 
$$\limsup_{k\to\infty}(T_\m^k J)(x)=J(1)>0=J_\m(1).$$
\nitem{(c)} For $a<0$, the cycle $(1,1)$ has negative length, and $\m$ is $R(X)$-regular because $J_\m(1)=0$, and we have $J_\m\in R(X)$, $J_\m(1)=\max\,[0,\,a+J_\m(1)]=(T_\m J_\m)(1)$, and for all $J\in R(X)$,
$$\limsup_{k\to\infty}(T^k_\m J)(1)=0=J_\m(1).$$
\endexample

\topinsert
\centerline{\hskip0pc\includegraphics[width=2.0in]{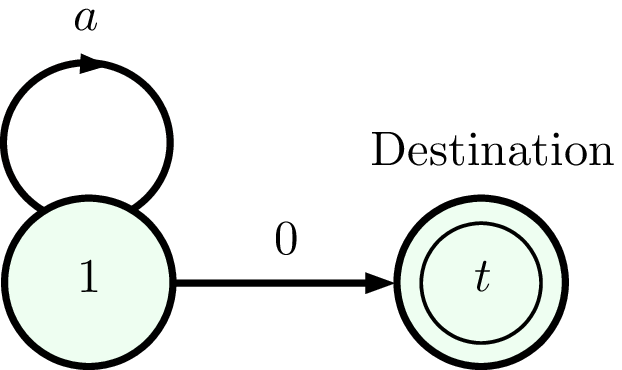}}
\vskip-0.5pc
\hskip-4pc\fig{0pc}{\figsimplrobusto.} {The subgraph of arcs ${\cal A}_\m$ corresponding to an improper policy $\m$, for the case of a single node 1 and a destination node $t$. The arcs lengths are shown in the figure.}
\endinsert

We now show one of our main results.

\xdef\propone{\propn}\propnum\show{myproposition}

\texshopbox{\proposition{\propone:}Let Assumption \assumptionone\ hold.
Then:
\nitem{(a)} The optimal cost function $\jstar$ of RSP is the unique fixed point of $T$ within $R(X)$.
\nitem{(b)} A policy  $\m^*$ is optimal for RSP if and only if
$T_{\m^*} \jstar =T\jstar$.  Moreover, there exists an optimal proper policy.
\nitem{(c)} We have $T^k J\to \jstar$ for all $J\in R(X)$.
\nitem{(d)} For any $J\in R(X)$, if $J\le TJ$ we have $J\le \jstar $, and if $J\ge TJ$ we have $J\ge \jstar $.}

\proof We verify the parts (a)-(f) of Assumption \assumptiontoz\ with $S=R(X)$. The result then will be proved using Prop.\ \propabstracto. To this end we argue as follows:
\nitem{(1)} Part (a) is satisfied since $S=R(X)$.
\nitem{(2)} Part (b) is satisfied since by Assumption \assumptionone(a), there exists at least one proper policy, which by Prop.\ \propproperchar\ is $R(X)$-regular. Moreover, for each $R(X)$-regular policy $\m$, we have $J_\m\in R(X)$. Since the number of all policies is finite, it follows that $\hat J\in R(X)$.
\nitem{(3)} To show that part (c) is satisfied, note that since by Prop.\ \propproperchar\ every $R(X)$-irregular policy $\m$ must be improper,  it follows from Assumption \assumptionone(b) that  the subgraph of arcs ${\cal A}_\m$ contains a cycle of  positive length. By Prop.\ \proppropimprop(d), this implies that for each $J\in R(X)$, we have $\limsup_{k\to\infty}(T^k_\m J)(x)=\infty$ for at least one $x\in X$.
\nitem{(4)} Part (d) is satisfied since $U(x)$ is a finite set.
\nitem{(5)} Part (e) is satisfied since $X$ is finite and $T_\m$ is a continuous function mapping the finite-dimensional space $R(X)$ into itself.
\nitem{(6)} To show that part (f) is satisfied, we note that by applying Prop.\ \propabstractt\ with $S=R(X)$, we have that $\hat J$ is the unique fixed point of $T$ within $R(X)$. It follows that for each  $J\in R(X)$, there exists a sufficiently large scalar $r>0$ such that the function $J'$ given by 
$$J'=\hat J-re,\qquad \forall\ x\in X,\eqnum\show{oneo}$$
where $e$ is the unit function, $e(x)\equiv1$, satisfies $J'\le J$ as well as 
$$J'=\hat J-re=T\hat J-re\le T(\hat J-re)=TJ',\eqnum\show{oneo}$$
where the inequality holds in view of Eqs.\  \tmmapdspo\ and \tmapdspt, and the fact $r>0$. 
\smskip
\pn Thus all parts of Assumption \assumptiontoz\ with $S=R(X)$ are satisfied, and Prop.\ \propabstracto\ applies with $S=R(X)$. Since  under Assumption \assumptionone, improper policies are $R(X)$-irregular [cf.\ Prop.\ \proppropimprop(d)] and so cannot be optimal, the minimax formulation of Section 2 is equivalent to RSP, and the conclusions of Prop.\ \propabstracto\ are precisely the results we want to prove. \qed

The following variant of the two-node Example \exampleirrego\ illustrates what may happen in the absence of Assumption \assumptionone(b), when there may exist improper policies that involve a nonpositive length cycle.

\xdef\exampleirregt{\exampl}\examplnum\show{myexample}

\xdef \figsimplrobust{\figr}\figrnum\show{myfigure}

\beginexample{\exampleirregt:}Let $X=\{1\}$, and consider the improper policy $\m$ with $Y\big(1,\m(1)\big)=\{1,t\}$ and the proper policy $\ol \m$ with  $Y\big(1,\ol\m(1)\big)=\{t\}$ (cf.\ Fig.\ \figsimplrobust). Let 
$$g\big(1,\m(1),1\big)=a\le0, \qquad g\big(1,\m(1),t\big)=0,\qquad g\big(1,\ol\m(1),t\big)=1.$$
Then it can be seen that under both policies, the longest path from 1 to $t$ consists of the arc $(1,t)$. Thus, 
$$J_\m(1)=0,\qquad J_{\ol\m}(1)=1,$$
so the improper policy $\m$ is optimal for the minimax problem \optcost, and strictly dominates the proper policy $\ol \m$ (which is optimal for the RSP version of the problem). To explain what is happening here, we consider two different cases:
\nitem{(1)}  $a=0$: In this case, the optimal policy $\m$ is both improper and $R(X)$-irregular, but with $J_\m(1)<\infty$. Thus the conditions of both Props.\ \propabstracto\ and \propone\ do not hold because Assumptions \assumptiontoz(c) and Assumption \assumptionone(b) are violated.
\nitem{(2)}  $a<0$: In this case, $\m$ is improper but $R(X)$-regular, so there are no $R(X)$-irregular policies. Then all the conditions of Assumption \assumptiontoz\ are satisfied, and Prop.\ \propabstracto\ applies. Consistent with this proposition, there exists an optimal $R(X)$-regular policy (i.e., optimal over both proper and improper policies), which however is improper and hence not an optimal solution for RSP.
\endexample

\topinsert
\centerline{\hskip0pc\includegraphics[width=4.5in]{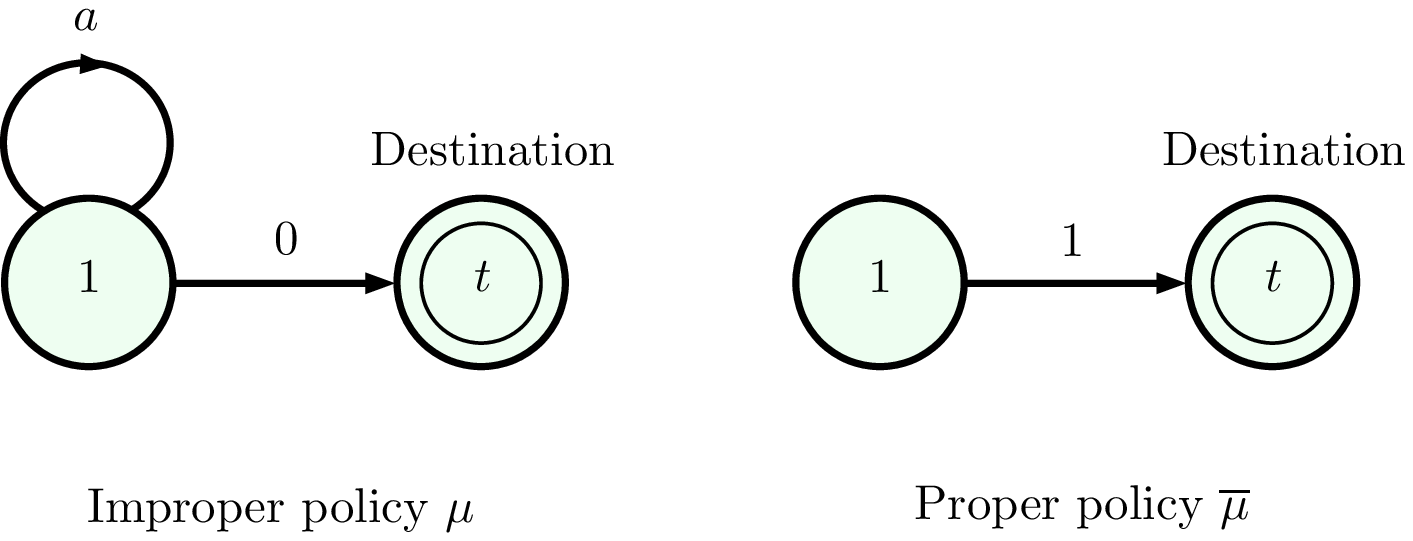}}
\vskip-1.pc
\hskip-4pc\fig{0pc}{\figsimplrobust.} {A counterexample involving a single node 1 in addition to the destination $t$. There are two policies, $\m$ and $\ol\m$, with corresponding subgraphs of arcs ${\cal A}_\m$ and ${\cal A}_{\ol\m}$, and arc lengths shown in the figure. The improper policy $\m$ is optimal when $a\le0$. It is $R(X)$-irregular if $a=0$, and it is $R(X)$-regular if $a<0$.}
\endinsert

We will next discuss modifications of Prop.\ \propone, which address the difficulties illustrated in the two cases of the preceding example.

\subsubsection{The Case  of Improper Policies with Negative Length Cycles}

\pn We note that by Prop.\ \propproperchar, the set of $R(X)$-regular policies includes not just proper policies, but also some improper ones (those $\m$ for which ${\cal A}_\m$ is destination-connected and all its cycles have negative length). As a result we can weaken  Assumption \assumptionone\ as long as it still implies Assumption \assumptiontoz\ so we can use Prop.\ \propabstracto\ to obtain corresponding versions of our main result of Prop.\ \propone. Here are two such weaker versions of Assumption \assumptionone.

\xdef\assumptiontwoo{\assumptionn}\assumptionnum\show{myproposition}

\texshopbox{\assumption{\assumptiontwoo:} Every policy $\m$ is either proper or else it is improper and its subgraph of arcs ${\cal A}_\m$ is destination-connected with all cycles having negative length.}

From Prop.\ \propproperchar, it follows that the preceding assumption is equivalent to all policies being $R(X)$-regular. The next assumption is weaker in that it allows policies $\m$ that are $R(X)$-irregular, as long as some cycle of ${\cal A}_\m$ has positive length.

\xdef\assumptiontwot{\assumptionn}\assumptionnum\show{myproposition}

\texshopbox{\assumption{\assumptiontwot:} \nitem{(a)} There exists at least one $R(X)$-regular policy.
\nitem{(b)} For every $R(X)$-irregular policy $\m$, some cycle in the subgraph of arcs ${\cal A}_\m$ has positive length.
}

Now by essentially repeating the proof of Prop.\ \propone, we see that Assumption \assumptiontwot\ implies Assumption \assumptiontoz, so that Prop.\ \propabstracto\ applies. Then we obtain the following variant of Prop.\ \propone.

\xdef\proptwo{\propn}\propnum\show{myproposition}

\texshopbox{\proposition{\proptwo:}Let either Assumption \assumptiontwoo\ or (more generally) Assumption \assumptiontwot\ hold.
Then:
\nitem{(a)} The function $\jstar$ of Eq.\ \jmimproper\ is the unique fixed point of $T$ within $R(X)$.
\nitem{(b)} A policy  $\m^*$ satisfies $J_{\m^*}=\jstar$, where $\jstar$ is the minimum of $J_\m$ over all $\m\in {\cal M}$ [cf.\ Eq.\ \optcost], if and only if
$T_{\m^*} \jstar =T\jstar$. Moreover, there exists an optimal $R(X)$-regular policy. 
\nitem{(c)} We have $T^k J\to \jstar$ for all $J\in R(X)$.
\nitem{(d)} For any $J\in R(X)$, if $J\le TJ$ we have $J\le \jstar $, and if $J\ge TJ$ we have $J\ge \jstar $.}

It is important to note that the optimal $R(X)$-regular policy $\m^*$ of part (b) above may not be proper, and hence needs to be checked to ensure that it solves the RSP problem (cf.\ Example \exampleirregt\ with $a<0$). Thus one would have to additionally prove that at least one of the optimal $R(X)$-regular policies is proper in order for the proposition to fully apply to RSP.

\subsubsection{The Case of Improper Policies with Zero Length Cycles}

\pn In some problems, it may be easier to guarantee nonnegativity rather than  positivity of the lengths of cycles corresponding to improper policies, which is required by Assumption \assumptionone(b). This is true for example in the important case where all arc lengths are nonnegative, i.e., $g(x,u,y)\ge0$ for all $x\in X$, $u\in U(x)$, and $y\in Y(x,u)$, as in case (1) of Example \exampleirregt.
Let us consider the following  relaxation of Assumption \assumptionone.

\xdef\assumptionthree{\assumptionn}\assumptionnum\show{myproposition}

\texshopbox{\assumption{\assumptionthree:}
\nitem{(a)} There exists at least one proper policy.
\nitem{(b)} For every improper policy $\m$, all cycles in the subgraph of arcs ${\cal A}_\m$ have nonnegative length.
}

Note that similar to the case of Assumption \assumptionone, we may guarantee that part (a) of the preceding assumption is satisfied by introducing a high cost termination action at each node. Then the policy that terminates at each state is proper.

For an analysis under the preceding assumption, we will use a perturbation approach that was introduced in Section 3.2.2 of [Ber13]. The idea is to consider a scalar $\d>0$ and a $\d$-perturbed problem, whereby each arc length $g(x,u,y)$ with $x\in X$ is replaced by $g(x,u,y)+\d$. As a result, a nonnegative cycle length corresponding to an improper policy as per Assumption \assumptionthree(b) becomes strictly positive, so Assumption \assumptionone\ is satisfied for the $\d$-perturbed problem, and Prop.\ \propone\ applies. We thus see that $\jstar_\d$, the optimal cost function of the $\d$-perturbed problem, is the unique fixed point of the mapping $T_\d$ given by
$$(T_\d J)(x)=\min_{u\in U(x)}H_\d(x,u,J),\qquad x\in X,$$
where $H_\d(x,u,J)$ is given by 
$$H_\d(x,u,J)=H(x,u,J)+\d.$$
Moreover there exists an optimal proper policy $\m_\d$ for the $\d$-perturbed problem, which by Prop.\ \propone(b), satisfies the optimality equation
$$T_{\m_\d,\d}\jstar_\d=T_\d \jstar_\d,$$
where $T_{\m,\d}$ is the mapping that corresponds to a policy $\m$ in the $\d$-perturbed problem:
$$(T_{\m,\d} J)(x)=H_\d\big(x,\m(x),J\big),\qquad x\in X.$$

We have the following proposition.

\xdef\propfour{\propn}\propnum\show{myproposition}

\texshopboxnb{\proposition{\propfour:} Let Assumption \assumptionthree\ hold, and let
$\hat J$ be the optimal cost function over the proper policies only,
$$\hat J(x)=\min_{\m:\, \hbox{\eightpoint proper}}J_\m(x),\qquad x\in X.$$
Then:
\nitem{(a)}
$\hat J=\lim_{\d\downarrow0}\jstar_\d.$}\texshopboxnt{
\nitem{(b)} $\hat J$ is the unique fixed point of $T$ within the set $\big\{J\in R(X)\mid J\ge \hat J\big\}$.
\nitem{(c)} We have $T^kJ\to \hat J$ for every
 $J\in R(X)$ with $J\ge \hat J$.
\nitem{(d)} Let $\m$ be a proper policy. Then  $\m$ is optimal within the class of proper policies (i.e., $J_\m=\hat J$) if and only if $T_{\m} \hat J =T\hat J$.
\nitem{(e)} There exists $\ol \d>0$ such that for all $\d\in(0,\ol\d]$, an optimal policy for the $\d$-perturbed problem is an optimal proper policy for the original RSP.
}

\proof (a) For all $\d>0$, consider an optimal proper policy $\m_\d$ of the $\d$-perturbed problem, i.e., one with cost $J_{\m_\d,\d}= \jstar_\d$. We have
$$\hat J\le J_{\m_\d}\le J_{\m_\d,\d}= \jstar_\d\le J_{\m',\d}\le J_{\m'}+N\d,\qquad \forall\ \m': \hbox{proper},
$$
where $N$ is the number of nodes of $X$ (since an extra $\d$ cost is incurred in the $\d$-perturbed problem every time a path goes through a node $x\ne t$, and any path under a proper $\m'$ contains at most $N$ nodes $x\ne t$). By taking the limit as $\d\downarrow0$ and then the minimum over all $\m'$ that are proper, it follows that
$$\hat J\le \lim_{\d\downarrow0}\jstar_\d\le \hat J,$$
so $\lim_{\d\downarrow0}\jstar_\d=\hat J$.

\smskip
\pn (b) For all proper $\m$, we have
$J_\m=T_\m J_\m\ge T_\m\hat J\ge T\hat J.$
Taking minimum over proper $\m$, we obtain $\hat J\ge T\hat J$.
Conversely, for all $\d>0$ and $\m\in{\cal M}$, we have
$$\jstar_\d=T\jstar_\d+\d e\le T_\m \jstar_\d+\d e.$$
Taking limit as $\d\downarrow0$, and using  part (a), we obtain $\hat J\le T_\m\hat J$ for all $\m\in{\cal M}$. Taking minimum over $\m\in{\cal M}$, it follows that $\hat J\le T\hat J$.  Thus $\hat J$ is a fixed point of $T$. The uniqueness of $\hat J$ will follow once we prove part (c).

\smskip
\pn (c) For all $J\in R(X)$ with $J\ge \hat J$ and proper policies $\m$, we have by using the relation $\hat J=T\hat J$ just shown in part (b),
$$\hat J=\lim_{k\to\infty}T^k\hat J\le \lim_{k\to\infty}T^k J\le \lim_{k\to\infty}T_\m^k J=J_\m.$$
Taking the minimum over all proper $\m$, we obtain
$$\hat J\le \lim_{k\to\infty}T^k J\le \hat J,\qquad \forall\ J\ge \hat J.$$

\smskip
\pn (d) If $\m$ is a proper policy with $J_\m=\hat J$, we have
$\hat J=J_{\m}=T_{\m}J_{\m}=T_{\m}\hat J,$
so, using  also the relation $\hat J=T\hat J$ [cf.\ part (a)], we obtain $T_{\m}\hat J=T\hat J$.
 Conversely, if  $\m$  satisfies $T_{\m}\hat J=T\hat J$, then from part (a), we have $T_\m \hat J=\hat J$ and hence $\lim_{k\to\infty}T_\m^k\hat J = \hat J$. Since $\m$ is proper, we have $J_\m=\lim_{k\to\infty}T_\m^k\hat J$, so  $J_\m=\hat J$.
 \smskip
 \pn {(e)} For every proper policy $\m$ we have $\lim_{\d\downarrow0}J_{\m,\d}= J_\m$. Hence if a proper $\m$ is not optimal for RSP, it is also nonoptimal for the $\d$-perturbed problem for all $\d\in [0,\d_\m]$, where $\d_\m$ is some positive scalar. Let $\ol \d$ be the minimum $\d_\m$ over the nonoptimal proper policies $\m$. Then for $\d\in(0,\ol\d]$, an optimal policy for the $\d$-perturbed problem cannot be nonoptimal for RSP. 
\qed 

Note that we may have $\jstar(x)<\hat J(x)$ for some $x$, but in RSP only proper policies are admissible, so by letting $\d\downarrow0$ we approach the optimal solution of interest. This happens for instance in Example \exampleirregt\ when $a=0$. For the same example $\hat J$ (not $\jstar$) can be obtained as the limit of $T^kJ$, starting from $J\ge \hat J$ [cf.\ part (c)]. The following example describes an interesting problem, where Prop.\  \propfour\ applies.

\old{Example \exampleirregt\ with $a<0$ also illustrates what may happen if the subgraph of arcs ${\cal A}_\m$ contains a cycle of negative length for some improper $\m$. Then $\m$ may be optimal with $J_\m$ real-valued, as in the case of Example \exampleirregt\ with $a<0$, or it may be optimal with $J_\m(x)=-\infty$, which would be true if the arc $(1,t)$ did not exist under $\m$ in Example \exampleirregt\ with $a<0$, or it may be strictly suboptimal as would be the case of Example \exampleirregt\ with $a<0$ but with the arc of the proper policy $\ol \m$ having negative length. }

\xdef\examplesearch{\exampl}\examplnum\show{myexample}

\beginexample{\examplesearch: (Minimax Search Problems)}Consider searching a graph with node set $X\cup \{t\}$, looking for an optimal node $x\in X$ at which to stop. At each  $x\in X$ we have two options: (1) stopping at a cost $s(x)$, which will stop the search by moving to  $t$, or (2) continuing the search by choosing a control $u\in U(x)$, in which case we will move to a node $y$, chosen from within a given set of nodes $Y(x,u)$ by an antagonistic opponent, at a cost $g(x,u,y)\ge0$.  Then Assumption \assumptionthree\ holds, since there exists a proper policy (the one that stops at every $x$). 

An interesting special case is when the stopping costs $s(x)$ are all nonnegative, while searching is cost-free [i.e., $g(x,u,y)\equiv 0$], but may lead in the future to nodes where a higher stopping cost will become inevitable. Then a policy that never stops is optimal but improper, but if we introduce a small perturbation $\d>0$ to the costs $g(x,u,y)$, we will make the lengths of all cycles positive, and Prop.\ \propfour\ may be used  to find an optimal policy within the class of proper policies. Note that this is an example where we are really interested in solving the RSP problem (where only the proper policies are admissible), and not its minimax version (where all policies are admissible).
\endexample
\vskip-1.5pc

\section{Computational Methods}

\pn We will now discuss computational methods that are patterned after the classical DP algorithms of value iteration and policy iteration  (VI and PI for short, respectively). In particular, the methods of this section are motivated by specialized stochastic shortest path algorithms.

\subsection{Value Iteration Algorithms}

\pn We have already shown as part of Prop.\ \propone\ that under Assumption \assumptionone, the VI algorithm, which sequentially generates $T^kJ$ for $k\ge0$, converges to the optimal cost function $\jstar$ for any starting function $J\in R(X)$.  We have also shown as part of Prop.\ \propfour\ that under Assumption \assumptionthree, the VI sequence $T^kJ$ for $k\ge0$, converges to $\hat J$, the optimal cost function over the proper policies only, for any starting function $J\ge \hat J$. We can extend these convergence properties to asynchronous versions of VI based on the monotonicity and fixed point properties of the mapping $T$. This has been known since the paper [Ber82] (see also [Ber83], [BeT89]), and we refer to the discussions in Sections 2.6.1, and 3.3.1 of [Ber13], which apply in their entirety when specialized to the RSP problem of this paper.

It turns out that for our problem, under Assumption \assumptionone\ or Assumption \assumptionthree,  the VI algorithm also terminates finitely when initialized with $J(x)=\infty$ for all $x\in X$ [it can be seen that in view of the form \tmapdspt\ of the mapping $T$, the VI algorithm is well-defined with this initialization]. In fact the number of iterations for termination is no more than $N$, where $N$ is the number of nodes in $X$, leading to polynomial complexity. This is consistent with a similar result for stochastic shortest path problems ([Ber12], Section 3.4.1), which relies on the  assumption of acyclicity of the graph of possible transitions under an optimal policy. Because this assumption is restrictive, finite termination of the VI algorithm is an exceptional property in stochastic shortest path problems. However, in the minimax case of this paper, an optimal policy $\m^*$ exists and is proper [cf.\ Prop.\ \propone(b) or Prop.\ \propfour(e)], so the graph of possible transitions under $\m^*$ is acyclic, and it turns out that finite termination of VI is guaranteed to occur. Note that in deterministic shortest path problems the initialization $J(x)=\infty$ for all $x\in X$, leads to polynomial complexity, and  generally works better in practice that other initializations (such as $J<\jstar$, for which the complexity is only pseudopolynomial, cf.\ [BeT89], Section 4.1, Prop.\ 1.2). 

 To show the finite termination property just described, let $\m^*$ be an optimal proper policy, consider the sets of nodes $X_0, X_1,\ldots$, defined
by 
$$X_0=\{t\},$$
$$X_{k+1}=\Big\{x\notin \cup_{m=0}^kX_m\mid y\in
\cup_{m=0}^kX_m\hbox{ for all }y\in Y\big(x,\m^*(x)\big)\Big\},\qquad k=0,1,\ldots,\xdef\xkodef{\lab}\eqnum\show{oneo}$$
and let $X_{\ol k}$ be the last of these sets that is nonempty.
Then in view of the acyclicity of the subgraph of arcs ${\cal A}_{\m^*}$, we have
$$\cup_{m=0}^{\ol k}X_m=X\cup \{t\}.$$
We will now show by induction that starting
from $J(x)\equiv\infty$ for all $x\in X$, the iterates $T^kJ$ of VI satisfy
$$(T^kJ)(x)= \jstar (x),\qquad \forall\ x\in \cup_{m=1}^kX_m,\ k=1,\ldots,\ol k.\xdef\exactconv{\lab}\eqnum\show{oneo}$$
Indeed, it can be seen that this is so for $k=1$. Assume that 
$(T^kJ)(x)= \jstar (x)$ if $x\in \cup_{m=1}^{k}X_m$.  Then, since $TJ\le J$ and $T$ is monotone, $(T^k J)(x)$ is monotonically nonincreasing, so that 
$$\jstar (x)\le  (T^{k+1}J)(x),\qquad \forall\ x\in X.\xdef\jstarineqo{\lab}\eqnum\show{oneo}$$
Moreover, by the induction hypothesis, the definition of the sets $X_k$, and
the optimality of $\m^*$, we have 
$$(T^{k+1}J)(x)\le H\big(x,\m^*(x),T^kJ\big)=
H\big(x,\m^*(x),\jstar\big)=\jstar (x),\qquad \forall\ x\in \cup_{m=1}^{k+1}X_m,\xdef\jstarineqt{\lab}\eqnum\show{oneo}$$
where the first equality follows from the form \tmmapdspo\ of $H$ and the fact that for all $x\in \cup_{m=1}^{k+1}X_m$, we have $y\in \cup_{m=1}^{k}X_m$ for all $y\in Y\big(x,\m^*(x)\big)$ by the definition \xkodef\ of $X_{k+1}$.
The two relations \jstarineqo\ and \jstarineqt\ complete the induction proof.

Thus under Assumption \assumptionone,  the VI method when started with $J(x)=\infty$ for all $x\in X$, 
will find the optimal costs of all the nodes in the set $\cup_{m=1}^kX_m$ after $k$ iterations; cf.\ Eq.\ \exactconv. The same is true under Assumption \assumptionthree, except that the method will find the corresponding optimal costs over the proper policies. In particular, all optimal costs
will be found after $\ol k\le N$ iterations, where $N$ is the number of nodes in $X$. This indicates that the behavior of the VI algorithm, when initialized with $J(x)=\infty$ for all $x\in X$, is similar to the one of the Bellman-Ford algorithm for deterministic shortest path problems. Still each iteration of the VI algorithm requires as many as $N$ applications of the mapping $T$ at every node. Thus it is likely  that the performance of the VI algorithm can be improved with a suitable choice of the initial function $J$, and with an asynchronous implementation that uses a favorable order of selecting nodes for iteration,  ``one-node-at-a-time" similar to the Gauss-Seidel method. This is consistent with the deterministic shortest path case, where there are VI-type algorithms, within the class of label-correcting methods, which are faster than the Bellman-Ford algorithm and even faster than efficient implementations of the Dijkstra algorithm for some types of problems; see e.g., [Ber98]. For the RSP problem, it can be seen that the best node selection order is based on the sets $X_k$ defined by Eq.\ \xkodef, i.e., iterate on the nodes in the set $X_1$, then on the nodes in $X_2$, and so on. In this case, only one iteration per node will be needed. While the sets $X_k$ are not known, an algorithm that tries to approximate the optimal order could be much more efficient that the standard ``all-nodes-at-once" VI method that computes the sequence $T^k J$, for $k\ge0$ (for an example of an algorithm of this type for stochastic shortest path problems, see  [PBT98]). The development of such more efficient VI algorithms is an interesting subject for further research, which, however, is beyond the scope of the present paper.

\xdef\examplevi{\exampl}\examplnum\show{myexample}
\xdef \figvidijkstraexample{\figr}\figrnum\show{myfigure}

\beginexample{\examplevi:}Let us illustrate the VI method for the problem of Fig.\ \figvidijkstraexample. The optimal policy is shown in this figure, and it is proper; this is consistent with the fact that Assumption \assumptionthree\ is satisfied. The table gives the iteration sequence of two VI methods, starting with $J_0=(\infty,\infty,\infty,\infty)$. The first method is the all-nodes-at-once method $J_k=T^kJ_0$, which finds $J^*$ in four iterations. In this example, we have $X_0=\{t\}$, $X_1=\{1\}$, $X_2=\{4\}$, $X_3=\{3\}$, $X_4=\{2\}$, and the assertion of Eq.\ \exactconv\ may be verified. The second method is the asynchronous VI method, which iterates one-node-at-a-time in the (most favorable) order 1, 4, 3, 2. The second method also finds $J^*$ in four iterations and with four times less computation.
\old{
An interesting additional observation is that if the lengths of the arcs (2,3) and (4,1) are changed to a value $\g<-1$, then the length of the two cycles of the graph becomes negative. As a result by Prop.\ \propproperchar, all policies become $R(X)$-regular, Assumption \assumptiontwoo\ holds, and by Prop.\ \proptwo, the VI method converges to $J^*$ starting from every $J\in R(X)$. This may be verified by the reader with a simple calculation. By contrast with the lengths of the arcs  (2,3) and (4,1) equal to $-1$, Assumptions  \assumptionone, \assumptiontwoo, and \assumptiontwot\ are violated and the VI convergence results of Props.\ \propone\ and \proptwo\ do not apply.
}
\endexample

\topinsert
\centerline{\hskip0pc\includegraphics[width=2.8in]{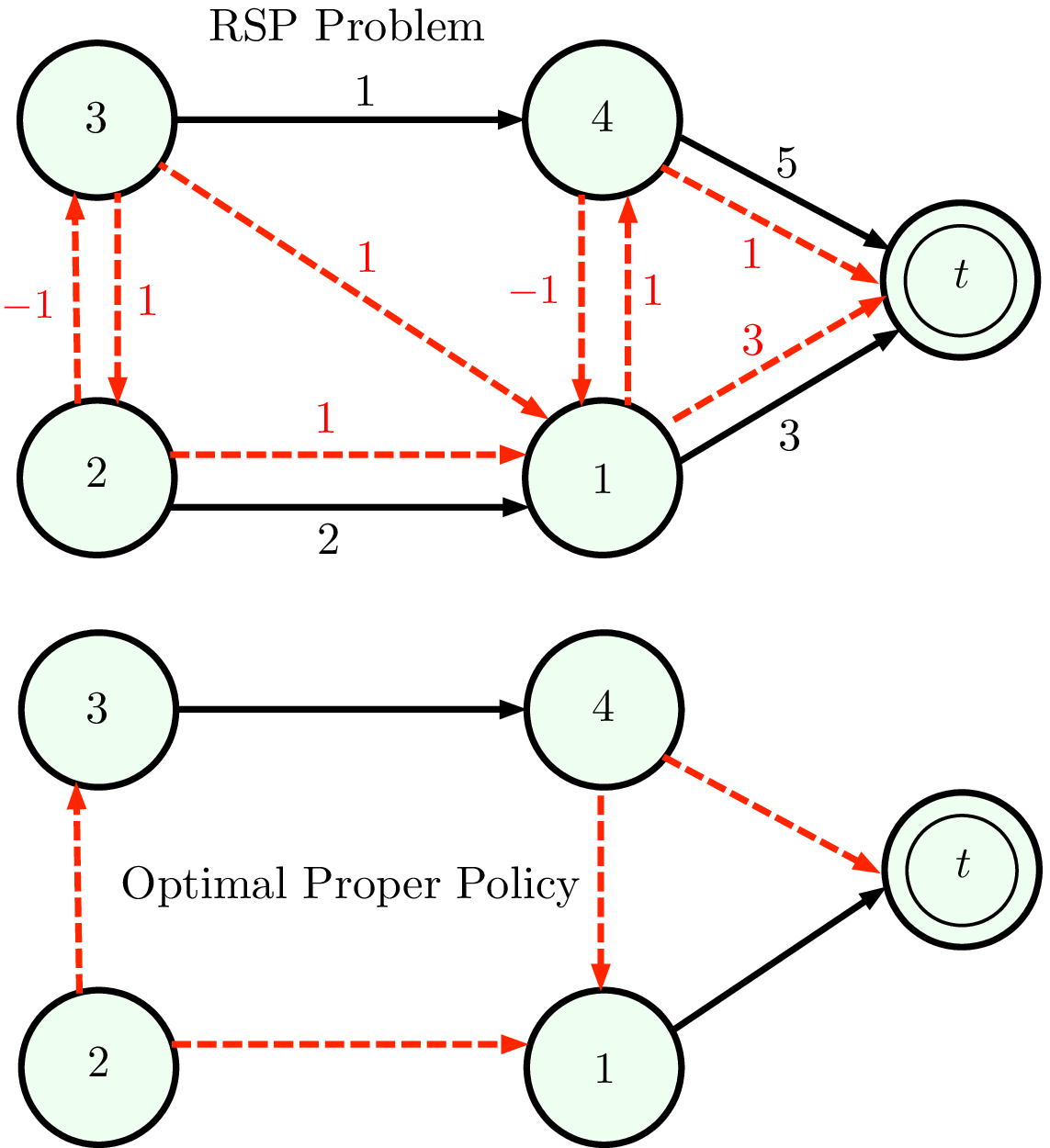}}
\vskip0.pc
\def\tablerule{\noalign{\hrule}}
\ninepoint$$\vbox{\offinterlineskip
\hrule
\halign{\vrule\hfill \ #\ \hfill &\vrule\hfill \ #\ \hfill 
&\vrule\hfill 
\ #\ \hfill \vrule\cr
&&\cr
{\bf Iteration \#\lower2ex\hbox{\ }\raise4ex\hbox{\ }}&\hbox{\bf
All-Nodes-at-Once VI Method}&\hbox{\bf One-Node-at-a-Time VI Method}\cr  \tablerule\cr
&&\cr
{0\lower2ex\hbox{\ }\raise4ex\hbox{\ }\hbox{\
}}&$J_0=(\infty,\infty,\infty,\infty)$&$J_0=(\infty,\infty,\infty,\infty)$\cr 
{1\lower2ex\hbox{\
}\raise2ex\hbox{\ }\hbox{\ }}&$J_1=TJ_0=(3,\infty,\infty,5)$&$J_1=(3,\infty,\infty,\infty)$\cr 
{2\lower2ex\hbox{\
}\raise2ex\hbox{\ }\hbox{\ }}&$J_2=T^2J_0=(3,5,6,2)$&$J_2=(3,\infty,\infty,2)$\cr 
{3\lower2ex\hbox{\
}\raise2ex\hbox{\ }\hbox{\ }}&$J_3=T^3J_0=(3,5,3,2)$&$J_3=(3,\infty,3,2)$\cr 
{4\lower2ex\hbox{\
}\raise2ex\hbox{\ }\hbox{\ }}&$J_4=T^4J_0=(3,4,3,2)=J^*$&$J_4=(3,4,3,2)=J^*$\cr
 \tablerule\cr}}$$  
\tenpoint
\fig{-6pt}{\figvidijkstraexample:}{An example RSP problem and its optimal policy. At each $X=\{1,2,3,4\}$ there are two controls: one (shown by a solid line) where $Y(x,u)$ consists of a single element, and another (shown by a broken line) where $Y(x,u)$ has two elements. Arc lengths are shown next to the arcs. Both the all-nodes-at-once and the one-node-at-a-time versions of VI terminate in four iterations, but the latter version requires four times less computation per iteration.
\vskip8pt}\endinsert

We finally note that in the absence of Assumption \assumptionone\ or Assumption \assumptiontwoo, it is possible that the VI sequence $\{T^kJ\}$ will not converge to $J^*$ starting from any $J$ with $J\ne J^*$. This can be seen with a simple deterministic shortest path problem involving a zero length cycle, a simpler version of Example \exampleirregt. Here there is a single node 1, aside from the destination $t$, and two choices at 1: stay at 1 at cost 0, and move to $t$ at cost 1. Then we have $J^*=0$, while $T$ is given by
$$TJ=\min\{J,\,1\}.$$
It can be seen that the set of fixed points of $T$ is $(-\infty,1]$, and contains $J^*$ in its interior. Starting with $J\ge1$, the VI sequence converges to 1 in a single step, while starting at $J\le 1$ it stays at $J$. This is consistent with Prop.\ \propfour(c), since in this example Assumption \assumptionthree\ holds, and we have $\hat J=1$. In the case of Example \exampleirregt\ with $a=0$, the situation is somewhat different but qualitatively similar. There it can be verified that $J^*=1$, the set of fixed points is $[0,1]$, $\{T^kJ\}$ will converge to 1 starting from $J\ge 1$,  will converge to 0 starting from $J\le 0$, and  will stay at $J$ starting from $J\in[0,1]$.

\subsection{Policy Iteration Algorithms}

\pn The development of PI algorithms for the RSP problem is straightforward given the connection with semicontractive models. Briefly, under  Assumption \assumptionone, based on the analysis of Section 3.3.2 of [Ber13], there are two types of PI algorithms. The first is a natural form of PI that generates proper policies exclusively. Let $\m_0$ be  an initial proper policy (there exists one by assumption). At the typical iteration $k$, we have a proper policy $\m_k$, and first compute $J_{\m_k}$ by solving a longest path problem over the corresponding acyclic subgraph of arcs ${\cal A}_{\m_k}$. We then compute a policy $\m_{k+1}$ such that 
$T_{\m_{k+1}} J_{\m_k}=TJ_{\m_k}$, by minimizing over $u\in U(x)$ the expression $H(x,u,J_{\m_k})$ of Eq.\ \tmmapdspo, for all $x\in X$.
We have
$$J_{\m_k}=T_{\m_{k}} J_{\m_k}\ge TJ_{\m_k}=T_{\m_{k+1}} J_{\m_k}\ge \lim_{m\to\infty}T_{\m_{k+1}}^m J_{\m_k}=J_{\m_{k+1}},\xdef\piineq{\lab}\eqnum\show{oneo}$$
where the second inequality follows from the monotonicity of $T_{\m_{k+1}}$, and the last equality is justified because $\m_k$ is proper and hence $ J_{\m_k}\in R(X)$,  so the next policy $\m_{k+1}$ cannot be improper [in view of Assumption \assumptionone(b) and Prop.\ \proppropimprop(d)]. In conclusion $\m_{k+1}$ must be proper and has improved cost over $\m_k$. 

Thus the sequence of policies $\{\m_k\}$ is well-defined and proper, and the corresponding  sequence $\{J_{\m_k}\}$ is nonincreasing. 
It then follows that $J_{\m_k}$ converges to $\jstar$ in a finite number of iterations. The reason is that from Eq.\ \piineq, we have that at the $k$th iteration, either strict improvement 
$$J_{\m_k}(x)> (TJ_{\m_k})(x)\ge J_{\m_{k+1}}(x)$$
is obtained for at least one node $x\in X$, or else $J_{\m_k}=T J_{\m_k}$, which implies that $J_{\m_k}=\jstar$ [since $\jstar$ is the unique fixed point of $T$ within $R(X)$, by Prop.\ \propone(a)] and $\m_k$ is an optimal proper policy.

Unfortunately, when there are improper policies, the preceding PI algorithm is somewhat limited, because an initial proper policy may not be known, and also because when asynchronous versions of the algorithm are implemented, it is difficult to guarantee that all the generated policies are proper. There is another algorithm, combining value and policy iterations, which has been developed in [BeY10], [BeY12], [YuB13a], [YuB13b] for a variety of DP models, including discounted, stochastic shortest path, and abstract, and is described  in Sections 2.6.3 and 3.3.2 of [Ber13]. This algorithm updates a cost function $J$ and a policy $\m$, but it also maintains an additional function $V$, which acts as a threshold to keep $J$ bounded and the algorithm convergent. The algorithm not only can tolerate the presence of improper policies, but can also be operated in asynchronous mode, whereby the value iterations, policy evaluation operations, and policy improvement iterations are performed one-node-at-a-time without any regularity. The algorithm is valid even in a distributed asynchronous environment, and in the presence of communication delays between processors. The specialization of this algorithm to RSP under Assumption \assumptionone\ is straightforward, and will be presented briefly in its asynchronous form, but without communication delays between processors.

We consider a distributed computing system with $m$ processors, denoted $1,\ldots,m$, a partition of the node set $X$ into sets $X_1,\ldots,X_m$, and an assignment of each subset $X_\ell$ to a processor $\ell\in\{1,\ldots,m\}$. The processors collectively maintain two functions $J_k(x)$ and $V_k(x)$, $x\in X$, and a policy $\m_k\in {\cal M}$, $k=0,1,\ldots$. We denote by $\min[V_k,J_k]$ the function in $E(X)$ that takes values $\min\big[V_k(x),\,J_k(x)\big]$ for $x\in X$. The initial conditions $J_0(x),V_0(x),\m_0(x)$, $x\in X$, are arbitrary. For each  processor $\ell$, there are two infinite disjoint subsets of times ${{\cal K}}_\ell,{\ol {\cal K}}_\ell \subset \{0,1,\ldots\}$, corresponding to local (within the subset $X_\ell$) policy improvement and policy evaluation iterations by that processor, respectively. More specifically, at each time $k$ and for each processor $\ell$, we have one of the following three possibilities:

\nitem{(a)}  {\it Local policy improvement\/}: If $k\in {\cal K}_\ell $, processor $\ell$ sets  for  all $x\in X_\ell $,
$$J _{k+1}(x)=V _{k+1}(x)=
\min_{u\in U(x)}H\big(x,u,\min[V_k,J_k] \big),\xdef \polimprb{\lab}\eqnum\show{oneo}$$
and sets $\m _{k+1}(x)$ to a $u$ that attains the above minimum.

\nitem{(b)}  {\it Local policy evaluation - Value iteration\/}:  If $k\in \ol {\cal K}_\ell $, processor $\ell$ sets  for  all $x\in X_\ell $,
$$J _{k+1}(x)=
H\big(x,\m _k(x),\min[V_k,J_k]\big),\xdef \polevalb{\lab}\eqnum\show{oneo}$$
and leaves $V $ and $\m $ unchanged, i.e., for  all $x\in X_\ell $, 
$V _{k+1}(x)=V _{k}(x)$, $\m _{k+1}(x)=\m _{k}(x).$
\smskip

\nitem{(c)} {\it No local change\/}: If $k\notin {{\cal K}}_\ell \cup {\ol {\cal K}}_\ell $, processor $\ell$ leaves $J $, $V $, and $\m $ unchanged, i.e., for  all $x\in X_\ell $, 
$$J _{k+1}(x)=J _{k}(x),\quad  V _{k+1}(x)=V _{k}(x),\quad \m _{k+1}(x)=\m _{k}(x).$$
\smskip
\pn In view of the form \tmmapdspo\ of the mapping $H$, the local policy improvement iteration \polimprb\ involves the solution of a static minimax problem, where the minimizing player chooses $u\in U(x)$ and the maximizing player chooses $y\in Y(x,u)$. The local policy evaluation iteration \polevalb\ involves a maximization over $y\in Y\big(x,\m_k(x)\big)$. 

The function $V_k$ in Eqs.\ \polimprb-\polevalb\ is reminiscent of a stopping cost in optimal stopping problems. The use of $V_k$ is essential for the asymptotic convergence of the algorithm to optimality, i.e., $J_k\to \jstar$, $V_k\to\jstar$, and for finite convergence of $\{\m_k\}$ to an optimal proper policy. Without $V_k$ the algorithm may potentially oscillate (there is an important counterexample that documents this type of phenomenon,  given in [WiB93]; see also the discussion in [BeY10], [BeY12], [Ber12]).

Note that the preceding algorithm includes as a special case a one-node-at-a-time asynchronous PI algorithm, whereby each node is viewed as a processor by itself, and at each iteration a single node is selected and a local policy improvement or local policy evaluation of the form \polimprb\ or \polevalb, respectively, is performed just at that node (see the discussion of Section 2.6 of [Ber12] or Section 2.6 of [Ber13]). This Gauss-Seidel type of algorithm is often considerably faster than all-nodes-at-once versions. The comparative evaluation of PI algorithms that use different initial conditions $J_0(x),V_0(x),\m_0(x)$, $x\in X$, and different orders of local policy improvement and policy evaluation iterations remains a subject for further research and experimentation.


\subsection{A Dijkstra-Like Algorithm for Nonnegative Arc Lengths}

\pn One of the most important computational approaches for the classical deterministic shortest path problem with nonnegative arc lengths is based on Dijkstra's algorithm, whereby the shortest distances of the nodes to the destination are determined  one-at-a-time in nondecreasing order. When properly implemented, this approach yields shortest path methods with excellent computational complexity and practical performance (see e.g., [AMO89], [Ber98]).

Dijkstra's algorithm has been extended to continuous-space shortest path problems in [Tsi95], and finds extensive application in large-scale computational problems involving the eikonal and other equations; see [Set99a], [Set99b]. For recent work in this area, see [ChV12], [AnV13], [CCV13], which give many other references. Dijkstra's algorithm has also been extended to finite-state stochastic shortest path problems, through the notion of a ``consistently improving optimal policy" (introduced in the 2001 2nd edition of the author's DP book, and also described in its 4th edition, [Ber12], Section 3.4.1). Roughly, with such a policy, from any node we may only go to a node of no greater optimal cost. While existence of a consistently improving optimal policy is a restrictive condition, the associated Dijkstra-like algorithm has found application in some special contexts, including large-scale continuous-space shortest path problems, where it is naturally satisfied; see [Vla08]. 
Our Dijkstra-like algorithm is patterned after the Dijkstra-like stochastic shortest path algorithm, but requires less restrictive conditions because an optimal proper policy has the essential character of a consistently improving policy when the arc lengths are nonnegative. As noted earlier, related Dijkstra-like algorithms were proposed by [GrJ08] (without the type of convergence analysis that we give here), and by [BaL15] (under the assumption that all arc lengths are strictly positive, and with an abbreviated convergence argument). We will assume the following:

\xdef\assumptiondijk{\assumptionn}\assumptionnum\show{myproposition}

\texshopbox{\assumption{\assumptiondijk:}
\nitem{(a)} There exists at least one proper policy.
\nitem{(b)} For every improper policy $\m$, all cycles in the subgraph of arcs ${\cal A}_\m$ have positive length.
\nitem{(c)} All arc lengths are nonnegative.
}

Parts (a) and (b) of the preceding assumption are just Assumption \assumptionone, under which the favorable results of Prop.\ \propone\ apply to RSP with both nonnegative and negative arc lengths. The arc length nonnnegativity assumption of part (c) provides additional structure, which provides the basis for the algorithm of this section.

Our Dijkstra-like algorithm maintains and updates a subset of nodes denoted $V$, and a number $J(x)$ for each $x\in X\cup\{t\}$, called the {\it label of $x$\/}. Initially,
$$V=\{t\},\qquad\qquad J(x)=\cases{0&if $x=t$,\cr
\infty&if $x\in X$.\cr}$$
At any given point in the algorithm, let $W$ be the set
$$W=\big\{x\mid J(x)<\infty,\,x\notin V\big\}.\xdef\setw{\lab}\eqnum\show{oneo}$$
The algorithm terminates when $V$ is empty. The typical iteration, assuming $V$ is nonempty, is as follows.

\texshopbox
{\pn {\bf Typical Iteration of the Dijkstra-Like Algorithm:}\smskip
\pn We remove  from $V$ a node $y^*$ such that
$$J(y^*)= \min_{y\in V}J(y),$$
and place it in $W$, i.e., replace $W$ with $W\cup\{y^*\}$. For every $x\notin W$, we let
$$\hat U(x)=\big\{u\in U(x)\mid Y(x,u)\subset W \hbox{ and }y^*\in Y(x,u)\big\},\eqnum\show{oneo}$$
and we update $J(x)$ and $V$ according to the following two cases:
\nitem{(1)} If $\hat U(x)$ is nonempty and $J(x)>\min_{u\in \hat U(x)}\max_{y\in Y(x,u)}\big[g(x,u,y)+J(y)\big]$, we set
 $$J(x)=\min_{u\in \hat U(x)}\max_{y\in Y(x,u)}\big[g(x,u,y)+J(y)\big],\xdef\jofxdef{\lab}\eqnum\show{oneo}$$
and we place $x$ in $V$ if it is not already there.
\nitem{(2)} Otherwise, we leave $J(x)$ and $V$ unchanged.
}
\smskip

Note that at each iteration of the preceding algorithm, the single node $y^*$ exits $V$, and enters the set $W$ of Eq.\ \setw. Thus $W$ is the set of nodes that have entered $V$ at some previous iteration but are not currently in $V$. 
Moreover, from the definition of the algorithm, once a node enters $W$ it stays in $W$ and never returns to $V$. Also, upon entrance of a node into $V$, its label changes from $\infty$ to some nonnegative number.  In the terminology of Dijkstra-like algorithms, $W$ is the set of nodes that are ``permanently labeled," and $V$ is the set of nodes that are ``candidates" for permanent labeling. 
We will show that all the nodes $x$ will enter $W$ in order of nondecreasing $J(x)$, and at the time of entrance, $J(x)=\jstar(x)$.

\xdef\propdijko{\propn}\propnum\show{myproposition}

\texshopbox
{\proposition{\propdijko:} Let Assumption \assumptiondijk\ hold. Then at the end of an iteration of the Dijkstra-like algorithm, we have $J(x')\ge J(x)$ for all $x'\notin W$ and $x\in
W$.
}

\proof We use  induction on
the iteration count. Clearly the assertion holds at the end of the initial iteration since then $W=\{t\}$, $J(t)=0$, and according to the formula \jofxdef\ for changing labels and the nonnegativity of the arc lengths, we have $J(x)\ge0$ for all $x\in X$.
Assume that the assertion holds for iteration $k-1$. Let $J(x)$ and $\tl J(x)$ denote the node labels at the start and the end of iteration $k$, respectively. Then by the minimum label rule for selection of $y^*$, we have
$$J(x')\ge J(y^*)\ge J(x)=\tl J(x),\qquad \forall\ x'\notin W\cup\{y^*\},\ x\in W\cup\{y^*\},\xdef\updateineq{\lab}\eqnum\show{oneo}$$
where the equality holds because the labels of all $x\in W\cup\{y^*\}$ will not change in iteration $k$. During iteration $k$ the labels of nodes $x'\notin W\cup\{y^*\}$ will change, if $\hat U(x')\ne\emptyset$, according to Eq.\ \jofxdef, so that
$$\eqalignno{\tl J(x')&=\min\lf[J(x'),\min_{u\in \hat U(x')}\max_{y\in Y(x',u)}\big[g(x',u,y)+J(y)\big]\ri]\cr
&\ge \min\big[J(x'),\,J(y^*)\big]\cr
&\ge J(x)\cr
&= \tl J(x),\qquad \forall\ x'\notin W\cup\{y^*\},\ x\in W\cup\{y^*\},\cr}$$
where the  first inequality holds because $g(x',u,y^*)\ge0$, and $y^*\in Y(x',u)$ for all $u\in \hat U(x')$, and the second  inequality and second equality hold because of Eq.\ \updateineq. 
The induction proof is complete. \qed

Since no node will enter $V$ twice, while exactly one node exits $V$ at each iteration, the algorithm will terminate after no more than $N+1$ iterations, where $N$ is the number of nodes in $X$. The next proposition shows that $V$ will become empty after exactly $N+1$ iterations, at which time $W$ must necessarily be equal to $X\cup\{t\}$.

\xdef\propdijkt{\propn}\propnum\show{myproposition}

\texshopbox
{\proposition{\propdijkt:} Let Assumption \assumptiondijk\ hold. The Dijkstra-like algorithm will terminate after exactly $N+1$ iterations with $V=\emptyset$ and $W=X\cup\{t\}$.
}

\proof Assume the contrary, i.e., that the algorithm will terminate after a number of iterations $k<N+1$. Then upon termination, $W$ will have $k$ nodes, $V$ will be empty, and the set 
$$\ol V=\{x\in X\mid x\notin W\}$$
will have $N+1-k$ nodes and thus be nonempty. Let $\ol \m$ be the proper policy, which is assumed to exist by Assumption \assumptiondijk(a). For each $x\in \ol V$ we cannot have $Y\big(x,\ol \m(x)\big)\subset W$, since then $x$ would have entered $V$ prior to termination, according to the rules of the algorithm. Thus for each $x\in \ol V$, there exists a node $y\in Y\big(x,\ol \m(x)\big)$ with $y\in \ol V$. This implies that the subgraph of arcs ${\cal A}_{\ol\m}$ contains a cycle of nodes in $\ol V$, thus contradicting the properness of $\ol\m$. \qed

There still remains the question of whether the final node labels $J(x)$, obtained upon termination of the algorithm, are equal to the optimal costs $\jstar(x)$. This is shown in the following proposition.

\xdef\propdijkth{\propn}\propnum\show{myproposition}

\texshopbox
{\proposition{\propdijkth:} Let Assumption \assumptiondijk\ hold. Upon termination of the Dijkstra-like algorithm, we have  $J(x)=\jstar(x)$ for all $x\in X$.
}

\proof For $k=0,1,\ldots$, let $V_k$, $W_k$, $J_k(x)$ denote the sets $V$, $W$, and the labels $J(x)$ at the start of iteration $k$, respectively, and let $y_k^*$ denote the minimum label node to enter $W$ during iteration $k$.  Thus, we have $W_{k+1}=W_k\cup\{y_k^*\}$,
$$J_{k+1}(x)=J_k(x),\qquad\forall\ x\in W_{k+1},$$
and
$$J_{k+1}(x)=\cases{\min\lf[J_k(x),\,\min_{u\in \hat U_k(x)}\max_{y\in Y(x,u)}\big[g(x,u,y)+J_k(y)\big]\ri]&if $x\in V_{k+1}$,\cr
\infty&if $x\notin W_{k+1}\cup V_{k+1}$,\cr}\eqnum\show{oneo}$$
where
$$\hat U_k(x)=\big\{u\in U(x)\mid Y(x,u)\subset W_k \hbox{ and }y^*\in Y(x,u)\big\}.$$

For each $k$ consider the sets of policies 
$$M_k(x)=\big\{\m:\hbox{proper}\mid \hbox{the nodes of all paths $p\in P(x,\m)$ lie in $W_k\cup\{x\}$}\big\}.$$
Note that $M_{k+1}(x)\supset M_k(x)$ since $W_{k+1}\supset W_k$, and that from the rule for a node to enter $V$, we have 
$$M_k(x)=\emptyset\qquad \hbox{if and only if}\qquad x\notin W_k\cup V_k,\xdef\mkdefformula{\lab}\eqnum\show{oneo}$$
[the reason why $M_k(x)\ne \emptyset$ for all $x\in W_k\cup V_k$ is that for entrance of $x$ in $V$ at some iteration there must exist $u\in U(x)$ such that $Y(x,u)$ is a subset of $W\cup\{y^*\}$ at that iteration].

We will prove by induction that for all $x\in X\cup \{t\}$, we have
$$J_k(x)=\cases{\min_{\m\in M_k(x)}\max_{p\in P(x,\m)}L_p(\m)&if $x\in W_k\cup V_k$,\cr
\infty&if $x\notin W_k\cup V_k$.\cr}\xdef\jkformula{\lab}\eqnum\show{oneo}$$
[In words, we will show that at the start of iteration $k$, $J_k(x)$ is the shortest ``minimax" distance from $x$ to $t$, using proper policies, which generate paths that start at $x$ and go exclusively through $W_k$. The idea is that since nodes in $W_k$ have smaller labels than nodes not in $W_k$ and the arc lengths are nonnegative, it would not be optimal to use a path that moves in and out of $W_k$.]
Equation \jkformula\
will imply that upon termination, when $M_{N+1}(x)$ is equal to the set of all proper policies, we have
$$J_{N+1}(x)=\min_{\m:\hbox{\eightpoint proper}\,}\max_{p\in P(x,\m)}L_p(\m),\qquad \forall\ x\in X,$$
which will prove the proposition. As can be expected, the proof is based on generalizations of proof ideas relating to the ordinary Dijkstra algorithm for the classical deterministic shortest path problem. We will often use the fact that for all $x\in W_{k+1}\cup V_{k+1}$ and $\m\in M_{k+1}(x)$ we have
$$\max_{p\in P(x,\m)}L_p(x)=\max_{y\in Y(x,\m(x))}\lf[g\big(x,\m(x),y\big)+\max_{p'\in P(y,\m)}L_{p'}(\m)\ri].\xdef\bellongest{\lab}\eqnum\show{oneo}$$
This is just the optimality equation for the longest path problem associated with $\m$ and the subgraph of arcs $A_\m$.

Initially, for $k=0$, we have $W_0=\emptyset$, $V_0=\{t\}$, $J_0(t)=0$, $J_0(x)=\infty$ for $x\ne t$, so Eq.\ \jkformula\ holds. Assume that Eq.\ \jkformula\ holds for some $k$. We will show that it holds for $k+1$. 

For $x\notin W_{k+1}\cup V_{k+1}$, we have $J_{k+1}(x)=\infty$, since such $x$ have never entered $V$ so far, and therefore their label was never reduced from the initial value $\infty$. This proves Eq.\ \jkformula\ with $k$ replaced by $k+1$ and $x\notin W_{k+1}\cup V_{k+1}$.

For $x=y_k^*$, from the definition \mkdefformula, the set of policies $M_{k+1}(y_k^*)$ is equal to $M_k(y^*_k)$, so by using also the induction hypothesis and the fact $J_{k+1}(y_k^*)=J_k(y_k^*)$, it follows that 
$$J_{k+1}(y_k^*)=J_k(y_k^*)=\min_{\m\in M_{k}(y_k^*)}\max_{p\in P(y_k^*,\m)}L_p(\m)=\min_{\m\in M_{k+1}(y_k^*)}\max_{p\in P(y_k^*,\m)}L_p(\m).$$
This proves Eq.\ \jkformula\ with $k$ replaced by $k+1$ and $x=y_k^*$.

For $x\in W_{k}\cup V_{k+1}$, we  write 
$$\min_{\m\in M_{k+1}(x)}\max_{p\in P(x,\m)}L_p(\m)=\min[E_1,\,E_2],$$
where 
$$E_1=\min_{\m\in M_{k}(x)}\max_{p\in P(x,\m)}L_p(\m),$$
which is equal to $J_k(x)$ by the induction hypothesis, and
$$E_2=\min_{\m\in M_{k+1}(x)/M_{k}(x)}\max_{p\in P(x,\m)}L_p(\m).$$
[The set $M_{k+1}(x)/M_{k}(x)$ may be empty, so here and later we use the convention that the minimum over the empty set is equal to $\infty$.] Thus for $x\in W_{k}\cup V_{k+1}$, we have
$$\min_{\m\in M_{k+1}(x)}\max_{p\in P(x,\m)}L_p(\m)=\min\lf[J_k(x),\,\min_{\m\in M_{k+1}(x)/M_{k}(x)}\max_{p\in P(x,\m)}L_p(\m)\ri],\xdef\quant{\lab}\eqnum\show{oneo}$$
and we need to show that the right-hand side above is equal to $J_{k+1}(x)$. To estimate the second term of the right-hand side, we consider the two separate cases where $x\in W_k$ and $x\in V_{k+1}$.

Assume first that $x\in W_k$. Then for each $\m\in M_{k+1}(x)/M_{k}(x)$, the subgraph of arcs ${\cal A}_\m$ must contain $y_k^*$, so
$$\max_{p\in P(x,\m)}L_p(\m)\ge \max_{p'\in P(y_k^*,\m)}L_{p'}(y_k^*)\ge J_k(y_k^*)\ge J_k(x),$$
where the first inequality follows in view of the nonnegativity of the arc lengths, the second inequality follows because paths in $P(y_k^*,\m)$ go exclusively through $W_k$ and thus can also be generated by some policy in $M_k(y_k^*)$, and the last inequality follows since nodes enter $W$ in order of nondecreasing label (cf.\ Prop.\ \propdijko). Thus the expression \quant\ is equal to $J_k(x)$, and hence is also equal to $J_{k+1}(x)$ (since $x\in W_k$).
This proves Eq.\ \jkformula\ with $k$ replaced by $k+1$ and $x\in W_k$.

Assume next that  $x\in V_{k+1}$. Then to estimate the  term $\min_{\m\in M_{k+1}(x)/M_{k}(x)}\max_{p\in P(x,\m)}L_p(\m)$ in Eq.\ \quant, we write 
$$M_{k+1}(x)/M_k(x)=\tl M_{k+1}(x)\cup \hat M_{k+1}(x),$$
where
$$\tl M_{k+1}(x)=\big\{ \m\in M_{k+1}(x)/M_k(x)\mid y_k^*\notin Y(x,\m(x))\big\},$$
is the set of policies for which there exists a path $p\in P(x,\m)$ that passes through $y^*_k$ after more than one transition, and
$$\hat M_{k+1}(x)=\big\{ \m\in M_{k+1}(x)/M_k(x)\mid y_k^*\in Y(x,\m(x))\big\},$$
is the set of policies for which there exists a path $p\in P(x,\m)$ that moves to $y^*_k$ at the first transition.

For all $\m\in \tl M_{k+1}(x)$, we have using Eq.\ \bellongest,
$$\eqalign{\max_{p\in P(x,\m)}L_p(\m)&=\max_{y\in Y(x,\m(x))}\lf[g\big(x,\m(x),y\big)+\max_{p'\in P(y,\m)}L_{p'}(\m)\ri]\cr
&\ge \max_{y\in Y(x,\m(x))}\lf[g\big(x,\m(x),y\big)+\min_{\m'\in M_{k+1}(y)}\max_{p'\in P(y,\m')}L_{p'}(\m')\ri]\cr
&=\max_{y\in Y(x,\m(x))}\lf[g\big(x,\m(x),y\big)+J_{k+1}(y)\ri]\cr
&=\max_{y\in Y(x,\m(x))}\lf[g\big(x,\m(x),y\big)+J_{k}(y)\ri],\cr}$$
where the last equality holds because we have shown that $J_{k+1}(y)=J_k(y)$ for all $y\in W_k$. 
Therefore, since for all $\m\in \tl M_{k+1}(x)$ there exists a $\m'\in M_{k}(x)$ such that $\m(x)=\m'(x)$, we have by taking the minimum over $\m\in \tl M_{k+1}(x)$ in the preceding relation,
$$\min_{\m\in \tl M_{k+1}(x)}\max_{p\in P(x,\m)}L_p(\m)\ge \min_{\m'\in M_{k}(x)}\max_{y\in Y(x,\m'(x))}\big[g(x,\m'(x),y)+J_{k}(y)\big]=\min_{\m'\in M_{k}(x)}\max_{p\in P(x,\m')}L_p(\m')=J_k(x).\eqnum\show{oneo}$$

For all $\m\in \hat M_{k+1}(x)$, we have
$$\max_{p\in P(x,\m)}L_p(x)=\max_{y\in Y(x,\m(x))}\lf[g\big(x,\m(x),y\big)+\max_{p'\in P(y,\m)}L_{p'}(\m)\ri];$$
cf.\ Eq.\ \bellongest.
Moreover, for all  $\m\in \hat M_{k+1}(x)$, we have $\m(x)\in \hat U_k(x)$ by the definitions of $\hat M_{k+1}(x)$ and $\hat U_k(x)$. It follows that 
$$\eqalign{\min_{\m\in \hat M_{k+1}(x)}\max_{p\in P(x,\m)}L_p(\m)&=\min_{\m\in \hat M_{k+1}(x)}\max_{y\in Y(x,\m(x))}\lf[g(x,\m(x),y)+\max_{p'\in P(y,\m)}L_{p'}(\m)\ri]\cr
&=\min_{u\in\hat U_k(x) }\max_{y\in Y(x,u)}\lf[g(x,u,y)+\min_{\m\in \hat M_{k+1}(x)}\max_{p'\in P(y,\m)}L_{p'}(\m)\ri]\cr
&=\min_{u\in\hat U_k(x) }\max_{y\in Y(x,u)}\lf[g(x,u,y)+\min_{\m\in M_{k+1}(y)}\max_{p'\in P(y,\m)}L_{p'}(\m)\ri]\cr
&=\min_{u\in\hat U_k(x) }\max_{y\in Y(x,u)}\big[g(x,u,y)+J_{k+1}(y)\big]\cr
&=\min_{u\in\hat U_k(x) }\max_{y\in Y(x,u)}\big[g(x,u,y)+J_k(y)\big],\cr}\xdef\specialonet{\lab}\eqnum\show{oneo}$$
where the third equality holds because for $y\in Y\big(x,\m(x)\big)$, the collections of paths $P(y,\m)$ under policies $\m$ in $\hat M_{k+1}(x)$ and $M_{k+1}(y)$,  are identical, and the last equality holds because we have already shown that $J_{k+1}(y)=J_k(y)$ for all $y\in W_k\cup\{y_k^*\}$. Thus from Eqs.\ \quant-\specialonet\ we obtain
$$\min_{\m\in M_{k+1}(x)}\max_{p\in P(x,\m)}L_p(\m)=\min\lf[J_k(x),\,\min_{u\in \hat U_k(x)}\max_{y\in Y(x,u)}\big[g(x,u,y)+J_k(y)\big]\ri].$$
Combining this equation with the update formula \jofxdef\ for the node labels, we have
$$\min_{\m\in M_{k+1}(x)}\max_{p\in P(x,\m)}L_p(\m)=J_{k+1}(x),$$
thus  proving Eq.\ \jkformula\ with $k$ replaced by $k+1$ and $x\in V_{k+1}$. This completes the induction proof of Eq.\ \jkformula\ and concludes the proof.
\qed

Since the algorithm terminates in $N+1$ iterations, and each iteration requires at most $O(AL)$ operations, where $A$ is the number of arcs and $L$ is the number of elements of $U$,  the complexity of the algorithm is bounded by $O(NAL)$. This complexity estimate may potentially be improved with the use of efficient data structures of the type used in efficient implementations of Dijkstra's algorithm in deterministic shortest path problems to expedite the selection of the minimum label node [i.e., $y^*\in \argmin_{y\in V}J(y)$]. However, we have not investigated this issue. It is also unclear how the Dijkstra-like algorithm compares with the finitely terminating VI algorithm and other asynchronous VI and PI algorithms discussed  in Sections 5.1, 5.2.

\xdef\exampledijkstra{\exampl}\examplnum\show{myexample}
\xdef \figvidijkstraexamplet{\figr}\figrnum\show{myfigure}

\topinsert
\centerline{\hskip0pc\includegraphics[width=2.8in]{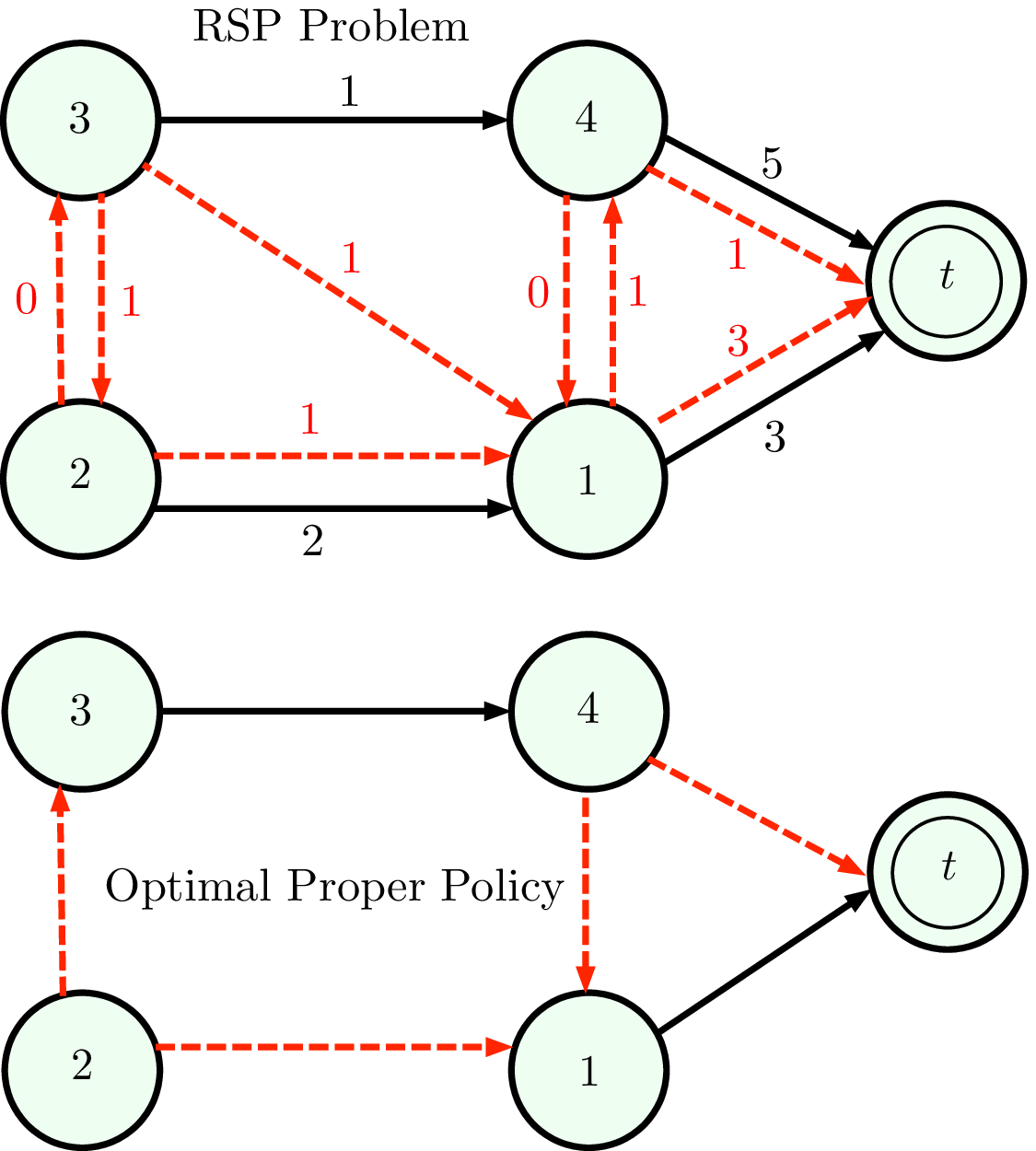}}
\vskip-0.7pc
\def\tablerule{\noalign{\hrule}}
\ninepoint$$\vbox{\offinterlineskip
\hrule
\halign{\vrule\hfill \ #\ \hfill &\vrule\hfill \ #\ \hfill 
&\vrule\hfill \ #\ \hfill &\vrule\hfill 
\ #\ \hfill \vrule\cr
&&\cr
{\bf Iteration \#\lower2ex\hbox{\ }\raise4ex\hbox{\ }}&\hbox{\bf
$V$ at Start of Iteration}&\hbox{\bf Labels at Start of Iteration} &\hbox{\bf Min. Label Node out of
$V$}\cr  \tablerule\cr
&&\cr
{0\lower2ex\hbox{\ }\raise4ex\hbox{\ }\hbox{\
}}&$\{t\}$&$(\infty,\infty,\infty,\infty,0)$&$t$\cr 
{1\lower2ex\hbox{\
}\raise2ex\hbox{\ }\hbox{\ }}&$\{1,4\}$&$(3,\infty,\infty,5,0)$&1\cr 
{2\lower2ex\hbox{\
}\raise2ex\hbox{\ }\hbox{\ }}&$\{2,4\}$&$(3,5,\infty,3,0)$&4\cr 
{3\lower2ex\hbox{\
}\raise2ex\hbox{\ }\hbox{\ }}&$\{2,3\}$&$(3,5,4,3,0)$&3\cr 
{4\lower2ex\hbox{\
}\raise2ex\hbox{\ }\hbox{\ }}&$\{2\}$&$(3,4,4,3,0)$&2\cr
\tablerule\cr}}$$  
\tenpoint
\fig{-16pt}{\figvidijkstraexamplet:}{The iterations of the Dijkstra-like algorithm for the RSP problem of Example \examplevi. The nodes exit $V$ and enter $W$ in the order 1, 4 3, 2.}\endinsert

\beginexample{\exampledijkstra:}We illustrate the Dijkstra-like algorithm with the RSP problem shown in Fig.\ \figvidijkstraexamplet. The table gives the iterations of the algorithm, and the results are consistent with Props.\ \propdijko-\propdijkth, as the reader may verify.
\endexample

\vskip-0.3pc

\subsection{Approximate Solution by Rollout}

\pn Finally let us consider algorithms with approximations. While we have assumed a finite number of nodes, there are many problems of practical interest where the number of nodes is extremely large, and the preceding algorithms are very time-consuming. This is particularly so in minimax control problems with imperfect state information, which are reformulated as problems of perfect state information using the sufficiently informative function approach of [Ber71], [BeR73]. In such cases one may consider minimax analogs of approximate DP or reinforcement learning approaches (see e.g., the books [BeT96], [SuB98], [Pow11], [Ber12]). The development of such methods is an interesting subject for further research. In what follows, we discuss the possibility for approximate solution using a rollout algorithm, which parallels related algorithms for finite horizon and stochastic shortest path problems (see e.g., [Ber05], [Ber12]).

The starting point of the rollout algorithm is a special proper policy $\m$, called the {\it base policy\/}. We  define the {\it rollout policy\/}, denoted $\ol \m$, as follows: 
for each $u\in U(x)$ and each $y\in Y(x,u)$, we compute $J_\m(y)$, and we set
$$\ol \m(x)\in \arg\min_{u\in U(x)}\max_{y\in Y(x,u)}\big\{g(x,u,y)+\tl J_\m(y)\big\},\qquad x\in X,\xdef\rollouteq{\lab}
\eqnum\show{oneo}$$
where $\tl J_\m(y)$ is equal to $J_\m(y)$ for $y\in X$ and $\tl J_\m(t)=0$ [cf.\ Eq.\ \tmmapdspotz]. The computation of $\ol \m(x)$ can be done on-line, only for the nodes $x$ that are encountered in the process of control. Moreover, assuming that $\tl J_\m(y)$ is precomputed for all $y$, and that the sets $U(x)$ and $Y(x,u)$ have a relatively small number of elements, the computation in Eq.\ \rollouteq\ can be performed quite fast. The same is true if for any $u\in U(x)$, $\tl J_\m(y)$ can be efficiently computed on-line for all $y\in Y(x,u)$. 

It can be seen that the rollout policy $\ol\m$ is just the policy obtained from the base policy $\m$ using a {\it single} policy iteration. In particular, under  Assumption \assumptionone,  the rollout policy improves on the base policy in the sense that
$$J_{\ol \m}(x)\le J_\m(x),\qquad \forall\ x\in X.$$
This is a well-known property of rollout algorithms for finite horizon and stochastic shortest path problems, and can be verified by writing for all $x\in X$
$$\eqalignno{(T_{\ol \m}J_\m)(x)&=(TJ_\m)(x)\cr
&=\min_{u\in U(x)}\max_{y\in Y(x,u)}\big\{g(x,u,y)+\tl J_\m(y)\big\}\cr
&\le \max_{y\in Y(x,\m(x))}\big\{g(x,\m(x),y)+\tl J_\m(y)\big\}\cr
&=(T_{\m}J_\m)(x)\cr
&=J_\m(x).\cr}\old{\eqnum\show{oneo}}$$
Applying $T_{\ol \m}$ repeatedly to both sides of the inequality $T_{\ol \m}J_\m\le J_\m$, we obtain [cf.\ Eq.\ \piineq] that $\ol\m$ is proper and that $J_{\ol \m}\le J_\m$. 

As an example of rollout algorithm, consider a pursuit-evasion problem with  state $x=(z_1,z_2)$, where $z_1$ is the location of the minimizer/pursuer and $z_2$  is the location of the maximizer/evader, in a two- or three-dimensional space. Then a suitable base policy $\m$ is for the pursuer is to follow a shortest path from $z_1$ to $z_2$ under the assumption that the evader will stay at his current location $z_2$ at all future times. To do this for all $(z_1,z_2)$ requires the solution of an all-pairs shortest path problem, which is possible in $O(N^3)$ time using the Floyd-Warshall algorithm [AMO89], [Ber98], where $N$ is the number of possible values of $z_1$ and $z_2$. Suppose that we have precomputed $\m(x)$ for all $x=(z_1,z_2)$ with this shortest path computation. Then the maximization 
$$\max_{y\in Y(x,u)}\big\{g(x,u,y)+\tl J_\m(y)\big\}$$
that is needed for the on-line computation of the rollout control $\ol \m(x)$ in Eq.\ \rollouteq\ requires the calculation of $J_\m(y)$ for each $y\in Y(x,u)$ with $y\ne t$. Knowing $\m$, each of these calculations is a tractable longest path computation in an acyclic graph of $N$ nodes. Note that the preceding algorithm can be adapted for the imperfect information case where the  pursuer knows $z_2$ imperfectly. This is possible by using a form of assumed certainty equivalence: the pursuer's base policy and the evader's maximization can be computed by using an estimate of the current location $z_2$ instead of the unknown true location. 

In the preceding pursuit-evasion example, the choice of the base policy was facilitated by the special structure of the problem. Generally, however, finding a suitable base policy whose cost function $J_\m$ can be conveniently computed is an important problem-dependent issue. We leave this issue as a subject for further research in the context of more specialized problems.
Finally, let us note that a rollout algorithm may be well-suited for  on-line suboptimal solution in cases where data may be changing or be revealed during the process of path construction.

\vskip-1pc

\section{Concluding Remarks and Further Research}

\pn We have considered shortest path problems with set membership uncertainty, and we have shown that they can be fruitfully analyzed in the context of abstract semicontractive models. We have thus proved the existence and uniqueness of the solution of Bellman's equation, and obtained conditions for optimality of a proper policy. Moreover, we have discussed the properties of algorithms of the value and policy iteration type, and we have proposed a finitely terminating Dijkstra-like algorithm for problems with nonnegative arc lengths. The comparative evaluation and the efficient implementation of these algorithms for specific types of applications, such as for example minimax search problems and pursuit-evasion, as well as  modifications to take advantage of special problem structures, is an interesting subject for further investigation.  

In this paper we have covered the important case of nonnegative arc lengths and improper policies with zero length cycles via the perturbation analysis of Section 4. However, there is an alternative line of analysis, which is based on the fact that when the arc lengths are nonnegative we have $T\bar J\ge\bar J$, bringing to bear the theory of monotone increasing DP models given in Chapter 4 of [Ber13], which embody the essential structure of negative DP  (see [Str66], or the texts [Put94], [Ber12]). This theory is somewhat different in character from the analysis of this paper.

Let us mention some interesting stochastic extensions of our RSP problem that involve an additional random variable at each stage.\footnote{\dag}{\ninepoint  This type of stochastic problem arises among others in a  context of discretization of the state space of a continuous-space minimax control problem, where randomization in the discretized problem's dynamics is introduced in order to reduce the error between the optimal cost function of the original continuous-space problem and  the optimal cost function of its discretized version (see [BSF94], [Tsi95], [PBT98], [Set99a], [Set99b]). There are also other stochastic shortest path-type formulations that involve at least in part a worst case viewpoint, through a risk-sensitive utility or constraints; see [DeR79], [Pat01], [EGS12], [CaR14], [Ber16].} In one extension of this type, when at node $x$, we choose control $u\in U(x)$, then a value of a random variable $z$ is selected from the finite set $\{1,\ldots,n\}$ with probabilities $p_{xz}(u)$, and then the successor node is chosen by an antagonistic opponent from a set $Y_z(x,u)\subset X\cup \{t\}$. To analyze this problem using a semicontractive model, the mapping $H$ of Eq.\ \tmmapdspo\ should be replaced by
$$H(x,u,J)=
\sum_{z=1}^np_{xz}(u)\max_{y\in Y_z(x,u)}\big[g(x,u,y)+\tl J(y)\big],\xdef\gendiscrmap{\lab}\eqnum\show{oneo}$$
where 
$$\tl J(y)=\cases{J(y)&if $y\in X$,\cr
0&if $y=t$.\cr}$$
A formulation as an abstract DP problem is then possible with an appropriately modified mapping $T_\m$, similar to Section 2, and the semicontractive theory may be applied similar to Section 4. This analysis and the associated algorithms are, however, beyond our scope. Note that when $n=1$, from Eq.\ \gendiscrmap\ we obtain the mapping \tmmapdspo\ of the RSP problem of Sections 1 and 2, while when $Y_z(x,u)$ consists of a single node, we obtain the mapping associated with a standard finite-state stochastic shortest path problem (see e.g., [Pal67], [BeT91], [Ber12], [BeY16]). Thus the semicontractive model that is based on the mapping \gendiscrmap\ generalizes both of these shortest path problems. In the case where $g(x,u,y)\ge 0$ for all $x\in X$, $u\in U(x)$, $z\in \{1,\ldots,n\}$, and $y\in Y_z(x,u)$, we have $T\bar J\ge \bar J$, and the theory of monotone increasing models of Sections 4.3, 4.4 of [Ber13] can be used to provide a first layer of analysis without any further assumptions.

In another extension of RSP, which involves randomization, when at node $x$, we choose control $u\in U(x)$, then a variable $w$ is chosen by an antagonistic opponent from a set $W(x,u)$, and then a successor node $y\in X\cup \{t\}$ is chosen according to probabilities $p_{wy}$. Here, to apply the line of analysis of the present paper, the mapping $H$ of Eq.\ \tmmapdspo\ should be replaced by
$$H(x,u,J)=
\max_{w\in W(x,u)}\lf[g(x,u,w)+\sum_{y\in X\cup\{t\}}p_{wy}\tl J(y)\ri],$$
where 
$$\tl J(y)=\cases{J(y)&if $y\in X$,\cr
0&if $y=t$.\cr}\xdef\aggregateo{\lab}\eqnum\show{oneo}$$
A somewhat more complex variation is given by
$$H(x,u,J)=
\sum_{z=1}^n p'_{xz}\max_{w\in W_z(x,u)}\lf[g(z,u,w)+\sum_{y\in X\cup\{t\}}p_{wy}\tl J(y)\ri],\xdef\aggregatet{\lab}\eqnum\show{oneo}$$
where, for each $x\in X$, $z$ is  a random variable taking values in $\{1,\ldots,n\}$ with probabilities $p'_{xz}$ [the two models based on Eqs.\ \aggregateo\ and \aggregatet\ coincide if $z$ takes values in $X$ and $p'_{xx}=1$ for all $x\in X$].
The resulting models again combine elements of RSP and a standard finite-state stochastic shortest path problem. They may also be viewed as instances of a generalized aggregation model of the type introduced in Section 7.3.7 of [Ber12].\footnote{\dag}{\ninepoint  In the context of generalized aggregation, for the mapping of Eq.\ \aggregateo, we have a high-dimensional RSP problem, whose states are represented by $w$, and a lower dimensional ``aggregate" RSP problem, whose states are represented by the nodes $x$. Then the $p_{wy}$ play the role of aggregation probabilities in the terminology of [Ber12]. A similar interpretation may be given for the mapping of Eq.\ \aggregatet.} Again the semicontractive theory may be applied, but the corresponding analysis is a subject for further research.

\vskip -1pc

\section{References}

\def\ref{\vskip1pt\pn}

\def\refer{\ref}

\ref[AHK07] Alfaro, L., Henzinger, L.\ T., and Kupferman, O., 2007.\ ``Concurrent Reachability Games," Theoretical Computer Science, Vol.\ 386, pp.\ 188-217.

\ref[AMO89] Ahuja, R.\ K., Magnanti, T.\ L., and Orlin, J.\ B., 1989.\ ``Network
Flows,'' in Handbooks in Operations Research and Management Science, Vol.\ 1,
Optimization, Nemhauser, G.\ L., Rinnooy-Kan, A.\ H.\ G., and Todd M.\ J.\ (eds.),
North-Holland, Amsterdam, pp.\ 211-369.

\ref[AlM12] Alton, K., and Mitchell, I.\ M., 2012.\ ``An Ordered Upwind Method with Precomputed Stencil and Monotone Node Acceptance for Solving Static Hamilton-Jacobi Equations," 
J.\ of Scientific Computing, Vol.\ 51, pp.\ 313-348.

\ref[AnV14] 
Andrews, J., and  Vladimirsky, A., 2014.\ ``Deterministic Control of Randomly-Terminated Processes," Interfaces and Free Boundaries, Vol.\ 16, pp.\ 1-40.

\ref[BBC11]  Bertsimas, D., Brown, D.\ B., and Caramanis, C., 2011.\ ``Theory and Applications of Robust Optimization," SIAM Rev., Vol.\ 53, pp.\ 464-501.

\ref[BBF95] Bardi, M., Bottacin, S., and Falcone, M., 1995.\ ``Convergence of Discrete Schemes for Discontinuous Value Functions of Pursuit-Evasion Games," New trends in dynamic games and applications, Vol.\ 3, pp.\ 273-304, Annals Intern.\ Soc.\ Dynamic Games, Birkhauser, Boston.

\ref[BBH08] Bopardikar, S.\ D., Bullo, F., and Hespanha, J.\ P., 2008.\ ``On Discrete-Time Pursuit-Evasion Games With Sensing Limitations, IEEE Trans.\ on Robotics, Vol.\ 24, pp.\ 1429-1439.

\ref [BGM95] Bertsekas, D.\ P., Guerriero, F., and Musmanno, R., 1995.\
``Parallel Shortest Path Methods for Globally Optimal Trajectories," High
Performance Computing: Technology, Methods, and Applications, (J.\ Dongarra
et al., Eds.), Elsevier.

\ref[BGN09] Ben-Tal, A., El Ghaoui, L., and Nemirovski, A., 2009.\ Robust Optimization, Princeton University Press, Princeton, N.\ J.

\ref[BSF94] Bardi, M., Soravia, P., and Falcone, M., 1994.\ ``Fully Discrete Schemes for the Value Function of Pursuit-Evasion Games," New trends in dynamic games and applications, Vol.\ 1, pp.\ 89-105, Annals Intern.\ Soc.\ Dynamic Games, Birkhauser, Boston.

\ref[BaB91] Basar, T., and Bernhard, P., 1991.\ H-Infinity Optimal Control and Related Minimax Design Problems: A Dynamic Game Approach, Birkhauser, Boston.

\ref[BaL15] Bardi, M., and Lopez, J.\ P.\ M., 2015.\ ``A Dijkstra-Type Algorithm for Dynamic Games,"  Dynamic Games and Applications, Vol.\ 6, pp.\ 1-14.

\ref[BaS91] Bardi, M., and Soravia, P., 1991.\ ``Approximation of Differential Games of Pursuit-Evasion by Discrete-Time Games," Differential Games - Developments in Modeling and Computation (Espoo, 1990), pp.\ 131-143, Lecture Notes in Control and Information Sci., 156, Springer, Berlin.

\ref [BeR71] Bertsekas, D.\ P., and Rhodes, I.\ B., 1971.\ ``Recursive State
Estimation for a Set-Membership Description of the Uncertainty," IEEE Trans.\
Automatic Control, Vol.\ AC-16, pp.\ 117-128.

\ref[BeR73] Bertsekas, D.\ P., and Rhodes, I.\ B., 1973.\ ``Sufficiently Informative Functions and the Minimax Feedback Control of Uncertain Dynamic Systems," IEEE Trans.\ on Automatic Control, Vol.\ AC-18, pp. 117-124.

\ref [BeS78]  Bertsekas, D.\ P., and Shreve, S.\ E., 1978.\  Stochastic Optimal
Control:  The Discrete Time Case, Academic Press, N.\ Y.; may be downloaded from http://web.mit.edu/dimitrib/www/home.html

\ref[BeS03] Bertsimas, D., and Sim, M., 2003.\ ``Robust Discrete Optimization and Network Flows," 
Math.\ Programming,  Vol.\ 98, pp.\ 49-71. 

\refer [BeT89] Bertsekas, D.\ P., and Tsitsiklis, J.\ N., 1989.\ Parallel and
Distributed Computation: Numerical Methods, Prentice-Hall, Englewood Cliffs,
N.\ J; republished by Athena Scientific, Belmont, MA, 1997.

\ref [BeT91]  Bertsekas, D.\ P., and Tsitsiklis, J.\ N., 1991.\ ``An Analysis of
Stochastic Shortest Path Problems,"
Math.\ of OR, Vol.\ 16, pp.\ 580-595.

\refer [BeT96] Bertsekas, D.\ P., and Tsitsiklis, J.\ N., 1996.\ Neuro-Dynamic Programming, Athena Scientific, Belmont, MA.

\ref[BeY10] Bertsekas, D.\ P., and Yu, H., 2010.\ ``Asynchronous Distributed Policy Iteration in Dynamic Programming,"  Proc.\ of Allerton Conf.\ on Communication, Control and Computing,  Allerton Park, Ill, pp.\ 1368-1374.

\refer[BeY12] Bertsekas, D.\ P., and Yu, H., 2012.\ ``Q-Learning and Enhanced Policy Iteration in Discounted Dynamic Programming,"  Math.\ of OR, Vol.\ 37, pp.\ 66-94.

\ref[BeY16]  Bertsekas, D.\ P., and Yu, H., 2016.\ ``Stochastic Shortest Path Problems Under Weak Conditions," Lab. for Information and Decision Systems Report LIDS-P-2909, MIT.

\ref [Ber71] Bertsekas, D.\ P., 1971.\ ``Control of Uncertain Systems With a
Set-Member\-ship Description of the Uncertainty," Ph.D.\ Dissertation, 
Massachusetts
Institute of Technology, Cambridge, MA (available in scanned form from
the author's www site).

\ref [Ber77] Bertsekas, D.\ P., 1977.\  ``Monotone Mappings with Application in
Dynamic Programming," SIAM J.\ on Control and Optimization, Vol.\ 15, pp.\
438-464.

\ref [Ber82] Bertsekas, D.\ P., 1982.\  ``Distributed Dynamic Programming," IEEE
Trans.\ Aut.\  Control, Vol.\ AC-27, pp.\ 610-616.

\ref[Ber83] Bertsekas, D.\ P., 1983.\ ``Asynchronous Distributed Computation of Fixed
Points," Math. Programming, Vol.\ 27, pp.\ 107-120.

\ref [Ber98] Bertsekas, D.\ P., 1998.\  Network Optimization: Continuous and Discrete
Models, Athena Scientific, Belmont, MA.

\ref[Ber05] Bertsekas, D.\ P., 2005.\ Dynamic Programming and Optimal Control, Vol.\ I, 3rd Edition, Athena Scientific, Belmont, MA.

\ref[Ber12] Bertsekas, D.\ P., 2012.\ Dynamic Programming and Optimal Control, Vol.\ II, 4th Edition, Athena Scientific, Belmont, MA.

\ref[Ber13] Bertsekas, D.\ P., 2013.\ Abstract Dynamic Programming, Athena Scientific, Belmont, MA.

\old{\ref[Ber14] Bertsekas, D.\ P., 2014.\ ``Infinite-Space Shortest Path Problems and Semicontractive Dynamic Programming," Lab. for Information and Decision Systems Report, MIT.}

\ref[Ber16] Bertsekas, D.\ P., 2016.\ ``Affine Monotonic and Risk-Sensitive Models in Dynamic Programming," Lab.\ for Information and Decision Systems Report LIDS-3204, MIT.

\ref[BlM08] Blanchini, F., and Miani, S., 2008.\ Set-Theoretic Methods in Control, Birkhauser, Boston, MA.

\ref [Bla99] Blanchini, F., 1999.\ ``Set Invariance in Control -- A Survey," 
Automatica, Vol.\ 35, pp.\ 1747-1768.

\ref[Bon07] Bonet, B., 2007.\ ``On the Speed of Convergence of Value Iteration on Stochastic Shortest-Path Problems," Math.\ of Operations Research, Vol.\ 32, pp.\ 365-373.

\ref[CCV14] Clawson, Z., Chacon, A., and Vladimirsky, A., 2014.\ ``Causal Domain Restriction for Eikonal Equations," SIAM J.\ on Scientific Computing, Vol.\ 36, pp.\ A2478-A2505.

\ref[CaB04] Camacho, E.\ F., and Bordons, C., 2004.\ Model Predictive Control, 2nd
Edition, Springer-Verlag, New York, N.\ Y.

\ref[CaR14] Cavus, O., and Ruszczynski, A., 2014.\ ``Computational Methods for Risk-Averse Undiscounted Transient Markov Models," Operations Research, Vol.\ 62, pp.\ 401-417.

\ref[ChV12] Chacon, A., and Vladimirsky, A., 2012.\ ``Fast Two-scale Methods for Eikonal Equations,"
SIAM Journal on Scientific Computing, Vol.\ 34, pp.\  A547-A577. 

\ref[DeR79] Denardo, E.\ V., and Rothblum, U.\ G., 1979.\ ``Optimal Stopping, Exponential Utility, and Linear Programming," Math.\ Programming, Vol.\ 16, pp.\ 228-244.

\ref [Den67] Denardo, E.\ V., 1967.\  ``Contraction Mappings in the Theory Underlying
Dynamic Programming," SIAM Review, Vol.\ 9, pp.\ 165-177.

\ref [Der70] Derman, C., 1970.\ Finite State Markovian Decision Processes,
Academic Press, N.\ Y.

\ref[Dre69] Dreyfus, S.\ E., 1969.\ ``An Appraisal of Some Shortest-Path Algorithms," 
Vol.\ 17, pp.\ 395-412.

\ref[EGS12] Ermon, S., Gomes, C., Selman, B., and Vladimirsky, A., 2012.\ ``Probabilistic Planning with Non-Linear Utility Functions and Worst-Case Guarantees," Proc.\ of the 11th Intern.\ Conf.\ on Autonomous Agents and Multiagent Systems (AAMAS 2012), Valencia, Spain.

\ref[Fal87] Falcone, M., 1987.\ ``A Numerical Approach to the Infinite Horizon
Problem of Deterministic Control Theory," Appl.\ Math.\ Opt., Vol.\ 15, pp.\
1-13.

\ref [FiV96] Filar, J., and Vrieze, K., 1996.\ Competitive Markov Decision Processes, Springer,  N.\ Y.

\ref[GLL99] Guibas,  L.\ J., Latombe, J.-C., LaValle, S.\ M.,  D. Lin, and Motwani, R. 1999.\ ``A Visibility-Based Pursuit-Evasion Problem," Intern.\ J.\ of Computational Geometry and Applications, Vol.\ 9, pp.\ 471-493.

\ref [GaP88] Gallo, G., and Pallottino, S., 1988.\ ``Shortest Path Algorithms," Annals
of Operations Research, Vol.\ 7, pp.\ 3-79.

\ref[GoR85] Gonzalez, R., and Rofman, E., 1985.\ ``On Deterministic Control
Problems: An Approximation Procedure for the Optimal Cost, Parts I, II," SIAM
J.\ Control Optimization, Vol.\ 23, pp.\ 242-285.

\ref[GrJ08] Grune, L., and Junge, O., 2008.\ ``Global Optimal Control of Perturbed Systems," J.\ of Optimization Theory and Applications, Vol.\ 136, pp.\ 411-429.

\ref [HCP99]  Hernandez-Lerma, O., Carrasco, O., and Perez-Hernandez.\ 1999.\  ``Markov Control Processes with
the Expected Total Cost Criterion: Optimality, Stability, and Transient Models," Acta Appl.\ Math.,
Vol.\ 59, pp.\ 229-269.

\ref[HKR93]
Hu, T.\ C., Kahng, A.\ B., and  Robins, G., 1993.\ ``Optimal Robust Path Planning in General Environments," IEEE Trans.\ on Robotics and Automation, Vol.\ 9, pp.\ 775-784.

\ref [HiW05] Hinderer, K., and Waldmann, K.-H., 2005.\ ``Algorithms for Countable
State Markov Decision Models with an Absorbing Set," SIAM J.\ of
Control and Optimization, Vol.\ 43, pp.\ 2109-2131.

\ref[JaC06] James, H.\ W., and Collins, E.\ J., 2006.\ ``An Analysis of Transient Markov Decision Processes," J.\
Appl.\ Prob., Vol.\ 43, pp.\ 603-621.

\old{
\ref[Jaq76] Jaquette, S.\ C.\ 1976.\ ``A Utility Criterion for Markov Decision Processes," Management Science, Vol.\ 23, pp.\ 43-49.
}

\ref[Ker00] Kerrigan, E.\ C., 2000.\ Robust Constraint Satisfaction: Invariant Sets and Predictive Control, Ph.D.\ Dissertation, Univ.\ of Cambridge, England.

\ref [KuD92] Kushner, H.\ J., and Dupuis, P.\ G., 1992.\ Numerical Methods for
Stochastic Control Problems in Continuous Time, Springer-Verlag, N.\ Y.

\ref [KuV97] Kurzhanski, A., and Valyi, I., 1997.\  Ellipsoidal Calculus
for Estimation and Control, Birkhauser, Boston, MA.

\ref[LaV06] LaValle, S.\ M., 2006.\ Planning Algorithms, Cambridge University Press, N.\ Y.

\ref[MHG88] Meggido, N., Hakimi, S.\ L., Garey, M.\ R., and Johnson, D.\ S., 1988.\ ``The Complexity of Searching a Graph,"
J.\ of ACM, Vol.\ 35, pp.\ 18-44.

\ref [MRR00] Mayne, D.\ Q., Rawlings, J.\ B., Rao, C.\ V., and Scokaert, P.\ O.\ M.,
2000.\ ``Constrained Model Predictive Control: Stability and Optimality," Automatica,
Vol.\ 36, pp.\ 789-814.

\ref [Mac02] Maciejowski, J.\ M., 2002.\ Predictive Control with Constraints,
Addison-Wesley, Reading, MA.

\ref[Mir14] Mirebeau, J-M., 2014.\ ``Efficient Fast Marching with Finsler Metrics," Numerische Mathematik, Vol.\ 126, pp.\ 515-557.

\ref[MoG04]  Montemanni, R., and Gambardella, L.\ M., 2004.\ ``An Exact Algorithm for the Robust Shortest Path Problem with Interval Data," Computers and Operations Research, Vol.\ 31, pp.\ 1667-1680.

\ref [MoL99] Morari, M., and Lee, J.\ H., 1999.\ ``Model Predictive Control: Past,
Present, and Future," Computers and Chemical Engineering, Vol.\ 23, pp.\ 667-682.

\ref[PBT98] Polymenakos, L.\ C., Bertsekas, D.\ P., and Tsitsiklis, J.\ N.,
1998.\ ``Efficient Algorithms for Continuous-Space Shortest Path
Problems," IEEE Trans.\  on Aut.\  Control, Vol.\ 43, pp.\ 278-283.

\ref[PaB99] Patek, S.\ D., and Bertsekas, D.\ P., 1999.\ ``Stochastic Shortest Path
Games," SIAM J.\ on Control and Optimization, Vol.\ 36, pp.\ 804-824.

\ref[Pal67] Pallu de la Barriere, R., 1967.\ Optimal Control Theory, Saunders, Phila; republished by Dover, N. Y., 1980.

\ref[Par76] Parsons, T.\ D., 1976.\ ``Pursuit-Evasion in a Graph," in Theory and Applications of Graphs, Y.\ Alavi and
D.\ Lick, Eds., Springer-Verlag, Berlin, pp.\ 426-441.

\ref[Pat01] Patek, S.\ D., 2001.\ ``On Terminating Markov Decision Processes with a Risk Averse Objective Function," Automatica, Vol.\ 37, pp.\ 1379-1386.

\ref[Pli78] Pliska, S.\ R., 1978.\ ``On the Transient Case for Markov Decision Chains with General State Spaces,"
in Dynamic Programming and Its Applications, M.\ L.\ Puterman
(ed.), Academic Press, N.\ Y.

\ref [Pow11] Powell, W.\ B., 2011.\  Approximate Dynamic Programming: Solving the Curses of Dimensionality, 2nd edition, J.\ Wiley and Sons, Hoboken, N.\ J.

\ref [Put94] Puterman, M.\ L., 1994.\  Markov Decision Processes: Discrete Stochastic Dynamic Programming, J.\ Wiley, N.\ Y.

\ref[RaM09] Rawlings, J.\ B., and Mayne, D.\ Q., 2009.\ Model Predictive Control Theory and Design, Nob Hill Publishing, Madison, WI.

\ref[Roc84]  Rockafellar, R.\ T., 1984.\ Network Flows and Monotropic
Programming, Wiley-Interscience, N.\ Y.

\ref[Set99a] Sethian, J.\ A., 1999.\ Level Set Methods and Fast Marching Methods 
Evolving Interfaces in Computational Geometry, Fluid Mechanics, Computer Vision, and
Materials Science, Cambridge Press,  N.\ Y.

\ref[Set99b] Sethian, J.\ A., 1999.\ ``Fast Marching Methods," SIAM Review, Vol.\ 41, pp.\ 199-235.

\ref [Sha53] Shapley, L.\ S., 1953.\  ``Stochastic Games," Proc.\ Nat.\ Acad.\
Sci.\ U.S.A., Vol.\ 39.

\ref [Str66] Strauch, R., 1966.\  ``Negative Dynamic Programming," Ann.\ Math.\
Statist., Vol.\ 37, pp.\ 871-890.

\ref[SuB98] Sutton, R.\  S., and Barto, A.\ G., 1998.\ Reinforcement Learning, MIT
Press, Cambridge, MA.

\ref [Tsi95] Tsitsiklis, J.\ N., 1995.\  ``Efficient Algorithms for Globally
Optimal Trajectories,'' IEEE Trans.\ Automatic Control, Vol.\ AC-40,
pp.\ 1528-1538.

\ref[VKS02] Vidal, R., Kim, H.\ J., Shim, D.\ H., and Sastry, S., 2002.\ ``Probabilistic Pursuit-Evasion Games: Theory, Implementation, and Experimental Evaluation," IEEE Trans.\ on Robotics and Automation, Vol.\ 18, pp.\ 662-669.

\ref[Vla08] Vladimirsky, A., 2008.\ ``Label-Setting Methods for Multimode Stochastic Shortest Path Problems on Graphs," Math.\ of Operations Research, Vol.\ 33, pp.\ 821-838.

\ref [Whi82] Whittle, P., 1982.\  Optimization Over Time, Wiley, N.\ Y., Vol.\
1, 1982, Vol.\ 2, 1983.

\ref[WiB93] Williams, R.\ J., and Baird, L.\ C., 1993.\ ``Analysis of
Some Incremental Variants of Policy Iteration: First Steps Toward
Understanding Actor-Critic Learning Systems,'' Report NU-CCS-93-11,
College of Computer Science, Northeastern University,
Boston, MA. 

\ref [Wit66] Witsenhausen, H.\ S., 1966.\ ``Minimax Control of Uncertain Systems,"
Ph.D.\ Dissertation,  Massachusetts Institute of Technology, Cambridge, MA.

\ref[YuB13a] Yu, H., and Bertsekas, D.\ P., 2013.\ ``Q-Learning and Policy Iteration Algorithms for Stochastic Shortest Path Problems," Annals of Operations Research, Vol.\ 208, pp.\ 95-132.

\ref[YuB13b] Yu, H., and Bertsekas, D.\ P., 2013.\ ``A Mixed Value and Policy Iteration Method for Stochastic Control with Universally Measurable Policies," Lab. for Information and Decision Systems Report LIDS-P-2905, MIT, July 2013. 

\ref[YuY98] Yu, G., and Yang, J., 1998.\ ``On the Robust Shortest Path Problem," Computers and Operations Research, Vol.\ 25, pp.\ 457-468.

\ref[Yu11] Yu, H., 2011.\ ``Stochastic Shortest Path Games and Q-Learning,"  Lab.\ for Information and Decision Systems Report LIDS-P-2875, MIT.

\end